\documentclass[pre,tighten]{revtex4}

\usepackage{amssymb,amsmath}

\usepackage{epsfig}

\usepackage{graphicx}
\usepackage{color}

\begin{document}
\title{Mass transport of impurities in a moderately dense granular gas}
\author{Vicente Garz\'{o}\footnote[1]{Electronic address: vicenteg@unex.es;
URL: http://www.unex.es/eweb/fisteor/vicente/} and Francisco Vega Reyes\footnote[2]{Electronic address: fvega@unex.es}}
\affiliation{Departamento de F\'{\i}sica, Universidad de Extremadura, E-06071 Badajoz, Spain}
\begin{abstract}

Transport coefficients associated with the mass flux of impurities immersed in a
moderately dense granular gas of hard disks or spheres described by the inelastic
Enskog equation are obtained by means of the Chapman-Enskog expansion. The transport
coefficients are determined as the solutions of a set of coupled linear integral
equations recently derived for polydisperse granular mixtures [V. Garz\'o, J. W. Dufty
and C. M. Hrenya, Phys. Rev. E {\bf 76}, 031304 (2007)]. With the objective of
obtaining theoretical expressions for the transport coefficients that are sufficiently
accurate for highly inelastic collisions, we solve the above integral equations by
using the second Sonine approximation. As a complementary route, we numerically solve
by means of the direct simulation Monte Carlo method (DSMC) the inelastic Enskog
equation to get the kinetic diffusion coefficient $D_0$ for two and three dimensions.
We have observed in all our simulations that the disagreement, for arbitrarily large
inelasticity, in the values of both solutions (DSMC and second Sonine approximation) is
less than 4\%. Moreover, we show that the second Sonine approximation to $D_0$ yields a
dramatic improvement (up to 50\%) over the first Sonine approximation for impurity
particles lighter than the surrounding gas and in the range of large inelasticity. The
results reported in this paper are of direct application in important problems in
granular flows, such as segregation driven by gravity and a thermal gradient. We
analyze here the segregation criteria that result from our theoretical expressions of
the transport coefficients.
\end{abstract}

\date{\today}
\maketitle

\section{Introduction
\label{sec1}}

The theoretical basis for a hydrodynamic description of ordinary (elastic) gases is
well established at low density using the Boltzmann kinetic equation. However, for
moderately dense gases there is no accurate and practical generalization of the
Boltzmann equation except for the idealized hard sphere fluid. For this intermolecular
potential, the Enskog kinetic equation takes into account the dominant positional
corrections to the Boltzmann equation due to excluded volume effects but, like the
Boltzmann equation, neglects velocity correlations (molecular chaos assumption) among
particles which are about to collide \cite{F72}. The Enskog equation is the only
available theory at present for making explicit calculations of the transport
properties of moderately dense gases and in any case the molecular chaos assumption is
expected to fail only in much denser systems (solid volume fractions typically are
larger than 0.4) \cite{DB77}. The extension to mixtures requires a revision of the
original Enskog theory for thermodynamic consistency (revised Enskog theory, or RET)
\cite{BE73}, and its application to hydrodynamics and Navier-Stokes (NS) transport
coefficients was carried out more than twenty years ago \cite{LCK83}.

Early attempts \cite{JM89,Z95,AW98,WA99} to extend the study of L\'opez de Haro, Cohen
and Kincaid \cite{LCK83} to {\em inelastic} hard sphere mixtures were restricted to
nearly elastic systems. In this case, the effect of inelasticity in the collisions is
taken into account only by the presence of a sink term in the energy balance equation
and, as a consequence, the expressions of the NS transport coefficients are the same as
those obtained for elastic systems. Moreover, those early works also assume energy
equipartition and so the partial temperatures for each species are equal to the
granular temperature. However, as the dissipation increases, different species of a
granular mixture have different partial temperatures $T_i$ and consequently, the energy
equipartition is seriously broken down ($T_i\neq T$) \cite{GD99,MP99,SD06}. The failure
of energy equipartition in granular fluids has also been confirmed by computer
simulations \cite{computer} and observed in real experiments \cite{exp} of agitated
mixtures. Results in the literature show that the deviation from equipartition depends
on the size and mass ratios of the particles of each species and the coefficients of
restitution of the system.

A more general extension of the RET to inelastic collisions has been recently carried
out by Garz\'o, Dufty and Hrenya \cite{GDH07,GHD07}. This theory covers some of the
aspects not taken into account in previous works \cite{JM89,Z95,AW98,WA99} and extends
previous results derived for monodisperse dense systems \cite{GD99a,L05} and dilute
binary mixtures \cite{GD02}. Specifically,  (i) it goes beyond the weak dissipation
limit so that it is expected to be applicable to a wide range of coefficients of
restitution, (ii) it takes into account the non-equipartition of granular energy, and
(iii) it has been formulated for multicomponent systems without limits on the number of
components. Therefore, this theory \cite{GDH07} subsumes all previous analysis for both
ordinary and granular gases, which are recovered in the appropriate limits
\cite{LCK83,GD99a,L05,GD02}. Nevertheless, as in the elastic case \cite{LCK83},
although the results are exact in the first order of spatial gradients, the explicit
form of the NS transport coefficients requires to solve a set of linear integral
equations. The standard method to get the kinetic and collisional contributions to
transport coefficients and cooling rate consists of approximating the solutions to
these integral equations by Maxwellians (at different temperatures) times truncated
Sonine polynomial expansions. For simplicity, usually only the lowest Sonine polynomial
(first Sonine approximation) is retained \cite{GHD07}, and the results obtained from this approximation compare very well with Monte
Carlo simulations of the Enskog equation in the case of the shear viscosity coefficient
of a mixture heated by an external thermostat \cite{GM03}. However, exceptions to this good
agreement are extreme mass or size ratios and strong dissipation, although these
discrepancies could be mitigated in part if one considers higher-order terms in the
Sonine polynomial expansion, as previous studies in the dilute limit indicate
\cite{GM04}. In fact, recent works for monodisperse gases have shown that higher order terms in Sonine polynomial expansions become increasingly important in the range of moderate and strong inelasticities \cite{HOB00,BP06}, and for this reason it has been of interest to calculate transport coefficients with more refined Sonine approaches \cite{NBSG07,GVM09}. However, the above works have been mainly focused with low density granular gases and many of the problems of practical interest in granular gases lie in the range of moderate densities. For this reason, it is important to determine the degree of
accuracy of at least the first Sonine approximation compared to the second Sonine
approximation for dense granular gases. Therefore, by testing an eventual gain of
accuracy with higher order Sonine approximations, our results in the present work will
contribute to the debate in the literature on the validity of a hydrodynamic
description of granular gases \cite{Go03}. In addition, the range of high inelasticities
has growing interest in experimental works \cite{Clerc08,VU08}. Another motivation to improve the evaluation of the NS transport coefficients lies in the fact that the reference homogeneous cooling state (HCS) is known to suffer a clustering instability, with inter-cluster distance inversely proportional to inelasticity \cite{GZ99}. In this context, we believe that a more accurate description of the HCS in the range of mild and strong inelasticities may help to refine the understanding of this interesting instability.

Needless to say, the evaluation of the NS transport coefficients for a dense granular
mixture beyond the first Sonine approximation is quite intricate, due mainly to the
coupling among the different integral equations obeying the transport coefficients. We
will thus make a first approach to the problem by considering the simple situation of a
granular binary mixture where the concentration of one of the species (of mass $m_0$
and diameter $\sigma_0$) is very small (impurity or tracer limit). Moreover, the tracer
limit has been of much interest in recent literature, for example in granular
segregation problems \cite{JY02,TAH03,BRM05,G08}. In the case of a tracer immersed in a
dense  granular gas, and as in a previous study for dilute gases \cite{GM04}, one can
assume that (i) the state of the dense gas (excess component of mass $m$ and diameter
$\sigma$) is not affected by the presence of impurities or tracer particles and, (ii)
one can also neglect collisions among tracer particles in their corresponding kinetic
equation. As a consequence, the velocity distribution function $f$ of the gas verifies
a closed Enskog equation while the velocity distribution function $f_0$ of the tracer
particles obeys the linear Enskog-Lorentz equation, which greatly simplifies the
development of Chapman-Enskog theory.

Under the above conditions, since the pressure tensor and heat flux of the mixture (gas
plus impurities) is the same as that for the gas \cite{GD99a,L05}, the mass transport
of impurities ${\bf j}_0$ is the relevant flux of the tracer problem. To first order in
the spatial gradients, three transport coefficients are involved in the constitutive
equation for ${\bf j}_0$: the kinetic diffusion coefficient $D_0$, the mutual diffusion
coefficient $D$ and the thermal diffusion coefficient $D^T$. Thus, the mass flux ${\bf
j}_0$ has the form \cite{GDH07}
\begin{equation}
\label{1.1} {\bf j}_0=-\frac{m_0^2}{\rho}D_{0}\nabla n_0-\frac{mm_0}{\rho}D \nabla
n-\frac{\rho}{T}D^T\nabla T,
\end{equation}
where $\rho=mn$ is the  total mass density, $n_0$ is the number density of the
impurities, $n$ is the number density of the gas particles, and $T$ is the granular
temperature. Therefore, the main goal of this paper is to determine $D_0$, $D$ and
$D^T$ up to the second Sonine approximation in terms of the coefficients of restitution
for the impurity--gas ($\alpha_0$) and gas--gas ($\alpha$) collisions, the parameters
of the system (masses and sizes) and the solid volume fraction $\phi$ occupied by the
gas. The calculations are rather intricate and we have taken advantage of some previous
calculations performed in Ref.\ \cite{GHD07} for multicomponent systems. In particular,
a previous expression for the coefficient $D_0$ obtained in the second Sonine
approximation for a dilute gas ($\phi=0$) is recovered \cite{GM04}. Analogously to the
previous analysis of the shear viscosity coefficient \cite{GM03}, kinetic theory
predictions for the diffusion coefficient $D_0$ are compared with numerical solutions
of the Enskog equation by using the well-known direct simulation Monte Carlo (DSMC)
method \cite{B94}. In the simulations, the diffusion coefficient is computed from the
mean-square displacement of impurities immersed in a dense granular gas undergoing the
homogeneous cooling state \cite{GM04}. Although the problem is time-dependent, a
transformation to a convenient set of dimensionless time and space variables
\cite{BRCG00} allows one to get a stationary diffusion equation where the coefficient
$D_0$ can be measured in the hydrodynamic regime (times large compared with the
characteristic mean free time).

Finally, once the explicit expression of the transport coefficients associated with the
mass flux are at hand, a segregation criterion based on thermal diffusion is derived.
This criterion shows the transition between the well-known Brazil-nut effect (BNE) and
the reverse Brazil-nut effect (RBNE) by varying the different parameters of the system.
This study complements a previous analysis recently carried out by one of the authors
\cite{G08} for a driven (heated) dense gas. As expected, our results show that the form
of the phase-diagrams for the BNE/RBNE transition depends sensitively on the value of
gravity relative to the thermal gradient and so it is possible to switch between both
states for given values of the parameters of the system.

The plan of the paper is as follows. In Sec.\ \ref{sec2} we describe the problem we are
interested in and offer a short summary of the set of inelastic Enskog equations for
the gas and the impurities. Section \ref{sec3} deals with the application of the
Chapman-Enskog method \cite{CC70} to solve the Enskog-Lorentz equation and get the set
of coupled linear integral equations verifying the transport coefficients $D_0$, $D$
and $D^T$. Then, these integral equations are approximately solved up to the second
Sonine approximation. Some technical details of the calculations are given in
Appendices \ref{appC} and \ref{appA}. In Sec.\ \ref{sec4} we illustrate the dependence
of the transport coefficients on the parameters of the system and compare the
theoretical results for the coefficient $D_0$ obtained in the first and second Sonine
approximation with those obtained by means of Monte Carlo simulations of the
Enskog-Lorentz equation for disks ($d=2$) and spheres ($d=3$). Segregation by thermal
diffusion is studied in Sec.\ \ref{sec5} and the paper is closed in Sec.\ \ref{sec6}
with a brief discussion on the results derived.

\section{Description of the problem\label{sec2}}

Let us consider a binary mixture of inelastic particles with collision rules according
to the smooth hard sphere model. Our system is described by the revised Enskog kinetic
equation \cite{GS95,BDS97} and, as a starting point, we consider in this work the
special case where the concentration of one of the components (the tracer or intruder)
is very small compared to that of the other (solvent or excess) component. In this
limit, the state of the granular gas (the solvent) is not affected by the presence of
the tracer particles and also the mutual interactions among the tracer particles can be
neglected as compared with their interactions with the particles of the solvent. At a
kinetic theory level, this implies that the velocity distribution function of the
solvent ($f$) and of the tracer particles ($f_0$) obey respectively the closed
(nonlinear) Enskog equation and the (linear) Enskog-Lorentz equation. This is formally
equivalent to study an impurity or intruder in a dense granular gas, and this will be
the terminology used here. Since in the tracer limit the pressure tensor and the heat
flux of dense binary mixtures are the same as those of the pure excess component (in
the absence of the tracer), here we will be interested in the evaluation of the
transport coefficients defining the mass flux of the intruder.

Let us start by offering a short review on some basic aspects of the set of inelastic
Enskog equations for the gas and the intruder. The granular dense gas is composed by
smooth inelastic hard disks ($d=2$) or spheres ($d=3$) of mass $m$ and diameter
$\sigma$. The inelasticity of collisions among all pairs is accounted for by a {\em
constant} coefficient of normal restitution $\alpha$ ($0\leq \alpha \leq 1$) that only
affects the translational degrees of freedom of grains.  The granular gas is in the
presence of a gravitational field ${\bf g}=-g \hat{{\bf e}}_z$, where $g$ is a positive
constant and $\hat{{\bf e}}_z$ is the unit vector in the positive direction of the $z$
axis. At moderate densities, we assume that the time evolution of the one-particle
velocity distribution function of the gas $f({\bf r}, {\bf v},t)$ is given by the
Enskog equation \cite{GS95,BDS97}
\begin{equation}
\label{2.1} \left( \partial _{t}+{\bf v}\cdot\nabla+{\bf g} \cdot
\frac{\partial}{\partial {\bf v}} \right)f({\bf r}, {\bf v},t)=J[{\bf
v}|f(t),f(t)],
\end{equation}
where the Enskog collision operator $J[{\bf v}|f,f]$ is
\begin{eqnarray}
\label{2.2}  J\left[ \mathbf{r}_{1},\mathbf{v}_{1}\mid f(t), f(t)\right] &\equiv &\sigma^{d-1}\int
d\mathbf{v}_{2}\int d\widehat{{\boldsymbol\sigma}}\Theta ( \widehat{{\boldsymbol\sigma}}\cdot
\mathbf{g}_{12})(\widehat{{\boldsymbol\sigma}}
\cdot \mathbf{g}_{12})  \notag \\
&&\times \left[ \alpha^{-2}\chi\left( \mathbf{r}_{1},\mathbf{r}
_{1}-{\boldsymbol\sigma}\right) f(\mathbf{r} _{1},\mathbf{v}_{1}^{\prime \prime
};t)f(\mathbf{r}_{1}-{\boldsymbol\sigma}
,\mathbf{v}_{2}^{\prime \prime };t)\right.  \notag \\
&&\left. -\chi\left( \mathbf{r}_{1},\mathbf{r}_{1}+{\boldsymbol\sigma}\right)
f(\mathbf{r}_{1},\mathbf{v} _{1};t)f(\mathbf{r}_{1}+{\boldsymbol\sigma},\mathbf{v}_{2};t)\right] .
\label{2.6}
\end{eqnarray}
Here, $\sigma$ is the hard sphere diameter, $\widehat{\boldsymbol {\sigma}}$ is a unit
vector along their line of centers, $\Theta $ is the Heaviside step function, and ${\bf
g}_{12}={\bf v}_{1}-{\bf v}_{2}$ is the relative velocity. The primes on the velocities
denote the initial values $\{{\bf v}_{1}^{\prime\prime}, {\bf v}_{2}^{\prime \prime}\}$
that lead to $\{{\bf v}_{1},{\bf v}_{2}\}$ following a binary collision in the hard
sphere model:
\begin{equation}
\label{2.3} {\bf v}_{1}^{\prime\prime}={\bf v}_{1}-\frac{1}{2}\left(1+\alpha^{-1}\right)
(\widehat{\boldsymbol {\sigma}}\cdot {\bf g}_{12})\widehat{\boldsymbol {\sigma}}, \quad {\bf v}_{2}^{\prime\prime
}={\bf v}_{2}+\frac{1}{2}\left( 1+\alpha^{-1}\right) (\widehat{\boldsymbol {\sigma}}\cdot {\bf
g}_{12})\widehat{\boldsymbol {\sigma}}.
\end{equation}
The quantity $\chi\left( \mathbf{r}_{1},\mathbf{r}_{1}+{\boldsymbol\sigma}\mid
n(t)\right)$ is the pair correlation function at contact as a functional of the
nonequilibrium density field $n({\bf r}, t)$, where
\begin{equation}
\label{2.4} n({\bf r}, t)=\int \; d{\bf v}f({\bf r}, {\bf v},t).
\end{equation}
In addition, the flow velocity and the {\em granular} temperature are defined respectively as
\begin{equation}
\label{2.5} {\bf u}({\bf r}, t)=\frac{1}{n({\bf r}, t)}\int \; d{\bf v} {\bf v} f({\bf
r},{\bf v},t),
\end{equation}
\begin{equation}
\label{2.6.1} T({\bf r}, t)=\frac{m}{d n({\bf r}, t)}\int \; d{\bf v}\; V^2 f({\bf
r},{\bf v},t),
\end{equation}
where ${\bf V}({\bf r},t)\equiv {\bf v}-{\bf u}({\bf r}, t)$ is the peculiar velocity.
The macroscopic balance equations for number density $n$, momentum density $m{\bf u}$,
and energy density $\frac{d}{2}nT$ follow directly from Eq.\ ({\ref{2.1}) by
multiplying with $1$, $m{\bf v}$, and $\frac{1}{2}mv^2$ and integrating over ${\bf v}$:
\begin{equation}
\label{2.7} D_{t}n+n\nabla \cdot {\bf u}=0\;,
\end{equation}
\begin{equation}
\label{2.8} D_{t}{\bf u}+(mn)^{-1}\nabla \cdot {\sf P}={\bf g}\;,
\end{equation}
\begin{equation}
\label{2.9} D_{t}T+\frac{2}{dn}\left(\nabla \cdot {\bf q}+P_{ij}\nabla_j u_{i}\right) =-\zeta T\;,
\end{equation}
where $D_{t}=\partial _{t}+{\bf u}\cdot \nabla$ is the material time derivative. The
microscopic expressions for the pressure tensor ${\sf P}$, the heat flux ${\bf q}$, and
the cooling rate $\zeta$ in terms of the velocity distribution function $f$ can be
found in Refs.\ \cite{GD99a} and \cite{L05}. Their explicit forms will be omitted here
for brevity.

Let us suppose now that an impurity or intruder of mass $m_0$ and diameter $\sigma_0$
is added to the gas. As said before, the presence of the intruder does not have any
effect on the state of the gas, so that its velocity distribution function is still
determined by the Enskog equation (\ref{2.1}). In addition, the macroscopic flow
velocity and temperature for the mixture composed by the dense gas plus the intruder
are the same as those for the gas, namely they are given by Eqs.\ (\ref{2.5}) and
(\ref{2.6.1}), respectively.  Under these conditions, the velocity distribution
function $f_0({\bf r}, {\bf v},t)$ of the intruder satisfies the linear Enskog-Lorentz
equation
\begin{equation}
\label{2.13} \left(\frac{\partial}{\partial t}+{\bf v}\cdot \nabla +{\bf g}\cdot
\frac{\partial}{\partial {\bf v}}\right) f_0({\bf r}, {\bf v},t)=J_{0}[{\bf v}|f_0(t),f(t)],
\end{equation}
where the collision operator $J_{0}[{\bf v}|f_0(t),f(t)]$ is now
\begin{eqnarray}
\label{2.14} J_{0}[{\bf r}_1, {\bf v}_1|f_0(t),f(t)]&=&\overline{\sigma}^{d-1} \int d\mathbf{v}_{2}\int
d\widehat{{\boldsymbol\sigma}}\Theta ( \widehat{{\boldsymbol\sigma}}\cdot
\mathbf{g}_{12})(\widehat{{\boldsymbol\sigma}}
\cdot \mathbf{g}_{12})  \notag \\
&&\times \left[ \alpha_0^{-2}\chi_0\left( \mathbf{r}_{1},\mathbf{r} _{1}-\overline{{\boldsymbol\sigma}}\right)
f_0(\mathbf{r} _{1},\mathbf{v}_{1}^{\prime \prime
};t)f(\mathbf{r}_{1}-\overline{{\boldsymbol\sigma}}
,\mathbf{v}_{2}^{\prime \prime };t)\right.  \notag \\
&&\left. -\chi_0\left( \mathbf{r}_{1},\mathbf{r}_{1}+\overline{{\boldsymbol\sigma}}
\right) f_0(\mathbf{r}_{1},\mathbf{v}
_{1};t)f(\mathbf{r}_{1}+\overline{{\boldsymbol\sigma}},\mathbf{v}_{2};t)\right] .
\end{eqnarray}
Here, $\overline{{\boldsymbol\sigma}}=\overline{\sigma}\widehat{{\boldsymbol\sigma}}$,
$\overline{\sigma}=(\sigma_0+\sigma)/2$, $\alpha_0$ ($0\leq \alpha_0 \leq 1$) is the
coefficient of restitution for intruder-gas collisions, and $\chi_0$ is the pair
correlation function for intruder-gas pairs at contact. The precollisional velocities
are given by
\begin{equation}
{\bf v}_{1}^{\prime \prime}={\bf v}_{1}-\frac{m}{m_0+m}\left( 1+\alpha _{0}^{-1}\right)
(\widehat{{\boldsymbol {\sigma }}}\cdot {\bf
g}_{12})\widehat{{\boldsymbol {\sigma }}} ,\nonumber\\
\end{equation}
\begin{equation}
 {\bf v}_{2}^{\prime \prime}={\bf v}_{2}+\frac{m_0}{m_0+m}\left( 1+\alpha
_{0}^{-1}\right) (\widehat{{\boldsymbol {\sigma }}}\cdot {\bf g}_{12})\widehat{ \boldsymbol
{\sigma}}. \label{2.3bis}
\end{equation}
As shown in Ref.\ \cite{SD06}, the operator $J_{0}\left[{\bf v}|f_0,f\right]$ is the same as that of an {\em
elastic} impurity ($\alpha_0=1$) with an effective mass
\begin{equation}
\label{2.15.0} m_0^*=m_0+\frac{1-\alpha_0}{1+\alpha_0}(m_0+m).
\end{equation}

The number density for the intruder is
\begin{equation}
\label{2.15} n_0({\bf r},t)=\int\; d{\bf v}f_0({\bf r},{\bf v},t).
\end{equation}
The intruder may freely loose or gain momentum and energy in its interactions with the
particles of the gas and, therefore, these are not invariants of the collision operator
$J_{0}[{\bf v}|f_0,f]$. Only the number density $n_0$ is conserved, whose continuity
equation is directly obtained from Eq.\ (\ref{2.13})
\begin{equation}
\label{2.16} D_{t}n_0+n_0\nabla \cdot {\bf u}+\frac{\nabla \cdot {\bf j}_0}{m_0}=0\;,
\end{equation}
where ${\bf j}_0$ is the mass flux for the intruder, relative to the local flow ${\bf u}$,
\begin{equation}
{\bf j}_{0}=m_{0}\int d{\bf v}\,{\bf V}\,f_0({\bf r},{\bf v},t). \label{2.17}
\end{equation}
At a kinetic level, an interesting quantity is the local temperature of the intruder, defined as
\begin{equation}
\label{2.18} T_0({\bf r}, t)=\frac{m_0}{d n_0({\bf r}, t)}\int \; d{\bf v}\, V^2
f_0({\bf r},{\bf v},t).
\end{equation}
This quantity measures the mean kinetic energy of the intruder. As confirmed by
computer simulations \cite{computer}, experiments \cite{exp} and kinetic theory
calculations \cite{GD99}, the global temperature $T$ and the temperature of the
intruder $T_0$ are in general different, so that the granular energy per particle is
not equally distributed between both components of the system.

\section{Mass transport of impurities}
\label{sec3}

In order to compute the mass flux ${\bf j}_0$ of impurities to first order in the
spatial gradients, we solve the Enskog-Lorentz equation by means of the Chapman-Enskog
(CE) expansion \cite{CC70}. This method, nowadays extensively used and tested in a
variety of problems in the field of rapid granular flows \cite{D02}, assumes the
existence of a {\em normal} solution in which all the space and time dependence of
$f_0$ occurs through the hydrodynamic fields $n_0$, $n$, ${\bf u}$ and $T$. The CE
procedure generates the normal solution explicitly by means of an expansion in the
gradients of the fields:
\begin{equation}
\label{3.1} f_0=f_0^{(0)}+\epsilon f_0^{(1)}+\cdots,
\end{equation}
where $\epsilon$ is a formal parameter measuring the nonuniformity of the system. The
application of the CE method to the Enskog equation for polydisperse granular mixtures
has been carried out very recently \cite{GDH07,GHD07}. Using those results, we consider
here the tracer limit ($x_0\equiv n_0/n\to 0$) for the linear integral equations
defining the transport coefficients associated with the mass flux. The first-order
contribution ${\bf j}_0^{(1)}$ to the mass flux is given by Eq.\ (\ref{1.1}), where the
kinetic diffusion coefficient $D_{0}$, the mutual diffusion coefficient $D$, and the
thermal diffusion coefficient $D^T$ are defined as
\begin{equation}
D^{T}=-\frac{m_0}{\rho d}\int d\mathbf{v}\mathbf{V}\cdot
\boldsymbol{\mathcal{A}}_{0}\left( \mathbf{V}\right) , \label{3.3}
\end{equation}
\begin{equation}
D_{0}=-\frac{\rho}{m_{0}n_{0}d}\int d\mathbf{v}\mathbf{V}\cdot
\boldsymbol{\mathcal{B}}_{0}\left( \mathbf{V}\right) ,\label{3.4}
\end{equation}
\begin{equation}
D=-\frac{1}{d}\int d\mathbf{v}\mathbf{V}\cdot \boldsymbol{\mathcal{C}}_{0}\left(
\mathbf{V}\right). \label{3.5}
\end{equation}
The quantities $\boldsymbol{\mathcal{A}}_{0}(\mathbf{V})$,
$\boldsymbol{\mathcal{B}}_{0}\left( \mathbf{V}\right)$, and
$\boldsymbol{\mathcal{C}}_{0}\left( \mathbf{V}\right)$ are the solutions of the
following set of coupled linear integral equations \cite{GDH07} :
\begin{equation}
\frac{1}{2}\zeta ^{(0)}\frac{\partial}{\partial {\bf V}}\cdot \left( {\bf
V}\boldsymbol{\mathcal{A}}_{0}\right)-\frac{1}{2}\zeta ^{(0)} \boldsymbol{\mathcal{A}}_0
-J_{0}^{(0)}[\boldsymbol{\mathcal{A}}_{0},f^{(0)}]=\mathbf{A}_{0}+J_{0}^{(0)}[f_0^{(0)},\boldsymbol{\mathcal{A}}],
\label{3.6}
\end{equation}
\begin{equation}
\frac{1}{2}\zeta ^{(0)}\frac{\partial}{\partial {\bf V}}\cdot \left( {\bf V}\boldsymbol{\mathcal{B}}_{0}\right)
-J_{0}^{(0)}[\boldsymbol{\mathcal{B}}_{0},f^{(0)}]=\mathbf{B}_{0}, \label{3.7}
\end{equation}
\begin{equation}
\frac{1}{2}\zeta ^{(0)}\frac{\partial}{\partial {\bf V}}\cdot \left( {\bf
V}\boldsymbol{\mathcal{C}}_{0}\right)-n\frac{\partial \zeta^{(0)}}{\partial n}\boldsymbol{\mathcal{A}}_0
-J_{0}^{(0)}[\boldsymbol{\mathcal{C}}_{0},f^{(0)}]=\mathbf{C}_{0}+J_{0}^{(0)}[f_0^{(0)},\boldsymbol{\mathcal{C}}],
\label{3.8}
\end{equation}
where $\zeta^{(0)}$ is the cooling rate to zeroth order (local homogeneous cooling
state) and $J_{0}^{(0)}[X,Y]$ is the operator
\begin{equation}
\label{3.9} J_{0}^{(0)}\left[{\bf v}_{1}|X,Y\right]=\chi_0^{(0)}\overline{\sigma}^{d-1}
\int d\mathbf{v}_{2}\int d\widehat{{\boldsymbol\sigma}}\Theta (
\widehat{{\boldsymbol\sigma}}\cdot \mathbf{g}_{12})(\widehat{{\boldsymbol\sigma}} \cdot
\mathbf{g}_{12})\left[ \alpha_0^{-2} X(\mathbf{V}_{1}^{\prime \prime
})Y(\mathbf{V}_{2}^{\prime \prime })- X(\mathbf{V}_{1})Y(\mathbf{V}_{2})\right],
\end{equation}
where $\chi_0^{(0)}$ is the intruder-gas pair correlation function at zeroth order.
The inhomogeneous terms of the integral equations (\ref{3.6})--(\ref{3.8}) are defined by
\begin{equation}
A_{0,i}\left( \mathbf{V}\right)=\frac{1}{2} V_{i}\frac{\partial}{\partial {\bf V}}\cdot \left(
\mathbf{V}f_{0}^{(0)}\right) -\frac{p}{\rho}\frac{\partial}{\partial V_{i}}
f_{0}^{(0)}+\frac{1}{2}\mathcal{ K}_{0,i}\left[\frac{\partial}{\partial {\bf V}}\cdot \left( \mathbf{V}
f^{(0)}\right) \right] , \label{3.10}
\end{equation}
\begin{equation}
{\bf B}_{0}\left( \mathbf{V}\right) = -{\bf V} f_{0}^{(0)},  \label{3.11}
\end{equation}
\begin{equation}
\label{3.12} C_{0,i}\left( \mathbf{V}\right) =-m^{-1}\frac{\partial p}{\partial n}\frac{\partial}{\partial V_{i}}
f_{0}^{(0)}- \frac{(1+\omega)^{-d}}{\chi_0^{(0)}T}
\left(\frac{\partial \mu_0}{\partial \phi}\right)_{T,n_0}
\mathcal{K}_{0,i}\left[f^{(0)}\right].
\end{equation}
In Eqs.\ (\ref{3.10})--(\ref{3.12}), the pressure $p$ is given by
\begin{equation}
p=nT\left[1+2^{d-2}\chi^{(0)}\phi(1+\alpha)\right], \label{3.12.1}
\end{equation}
$\omega\equiv\sigma_0/\sigma$ is the size ratio and
$\mu_0$ is the chemical potential of the intruder. Furthermore,
\begin{equation}
\phi\equiv \frac{\pi^{d/2}}{2^{d-1}d\Gamma(d/2)} n\sigma^d \label{3.13.1}
\end{equation}
is the solid volume fraction and the operator $\mathcal{K}_{0,i}[X]$ is defined as
\begin{equation}
\label{3.14} \mathcal{K}_{0,i}[X] =\overline{\sigma}^{d}\chi _{0}^{(0)}\int d
\mathbf{v}_{2}\int d\widehat{\boldsymbol {\sigma }}\Theta (\widehat{\boldsymbol
{\sigma}} \cdot \mathbf{g}_{12})(\widehat{\boldsymbol {\sigma}}\cdot \mathbf{g}_{12})
\widehat{\sigma}_i \left[ \alpha _{0}^{-2}f_{0}^{(0)}(\mathbf{V}_{1}^{\prime
\prime})X(\mathbf{V}_{2}^{\prime \prime
})+f_{0}^{(0)}(\mathbf{V}_{1})X(\mathbf{V}_{2})\right] .
\end{equation}

Upon writing Eqs.\ (\ref{3.6})--(\ref{3.8}), use has been made of the expression of the
first-order distribution function $f^{(1)}$ of gas particles. Its form has been derived
in Refs.\ \cite{GD99a,L05} and reads
\begin{equation}
\label{3.15} f^{(1)}=\boldsymbol{\mathcal{A}}\cdot \nabla T+\boldsymbol{\mathcal{C}}\cdot
\nabla n+\boldsymbol{\mathcal{D}}:\nabla {\bf u}+E \nabla\cdot {\bf u},
\end{equation}
where the coefficients $\boldsymbol{\mathcal{A}}$, $\boldsymbol{\mathcal{C}}$,
$\boldsymbol{\mathcal{D}}$ and $E$ are functions of the peculiar velocity ${\bf V}$ and
the hydrodynamic fields.  According to Eqs.\ (\ref{3.6})--(\ref{3.8}), only the
coefficients $\boldsymbol{\mathcal{A}}$ and $\boldsymbol{\mathcal{C}}$ are involved in
the evaluation of the transport coefficients $D_0$, $D$ and $D^T$. The linear integral
equations verifying $\boldsymbol{\mathcal{A}}$ and $\boldsymbol{\mathcal{C}}$ as well
as their expressions up to  the second Sonine approximation are given in Appendix
\ref{appC}.

It is worthwhile to remark that the quantities $\boldsymbol{\mathcal{A}}_{0}$ and
$\boldsymbol{\mathcal{C}}_{0}$ associated with the intruder are coupled with their
corresponding counterparts $\boldsymbol{\mathcal{A}}$ and $\boldsymbol{\mathcal{C}}$ of
the host gas through the integral equations (\ref{3.6}) and (\ref{3.8}), respectively.
A direct consequence of this coupling is that the mass flux of the intruder (\ref{1.1})
inherits gradient terms ($\nabla n$ and $\nabla T$) from those of the autonomous host
equations. Moreover, the external field does not occur in the constitutive equation
(\ref{1.1}) for the mass flux. This is due to the particular form of the gravitational
force.

\section{Second Sonine polynomial approximation}
\label{sec4}

For practical purposes, the integral equations (\ref{3.6})--(\ref{3.8}) can be solved
by using a Sonine polynomial expansion. With the motivations explained in the
Introduction, our goal here is to determine the diffusion coefficients $D_0$ and $D$
and the thermal diffusion coefficient $D^T$ up to the second Sonine approximation. In
this case, the quantities $\boldsymbol{\mathcal{A}}_0$, $\boldsymbol{\mathcal{B}}_0$
and $\boldsymbol{\mathcal{C}}_0$ are approximated by
\begin{equation}
\label{4.1} \boldsymbol{\mathcal{A}}_{0}({\bf V})\to -f_{0,M}({\bf V})\left[\frac{\rho}{n_0T_0}{\bf
V}D^T+a_0{\bf S}_0({\bf V}) \right],
\end{equation}
\begin{equation}
\label{4.2} \boldsymbol{\mathcal{B}}_{0}({\bf V})\to -f_{0,M}({\bf V})\left[\frac{m_0^2}{\rho T_0}{\bf
V}D_{0}+b_0{\bf S}_0({\bf V}) \right] ,
\end{equation}
\begin{equation}
\label{4.3} \boldsymbol{\mathcal{C}}_{0}({\bf V})\to -f_{0,M}({\bf V})\left[\frac{m_0}{ n_0T_0}{\bf
V}D+c_0{\bf S}_0({\bf V}) \right] ,
\end{equation}
where
\begin{equation}
\label{4.4} {\bf S}_0({\bf V})=\left(\frac{1}{2}m_0V^2-\frac{d+2}{2}T_0\right){\bf V},
\end{equation}
and $f_{0,M}({\bf V})$ is a Maxwellian distribution at the temperature $T_0$ of the intruder, i.e.,
\begin{equation}
\label{4.4.1} f_{0,M}({\bf V})=n_0 \left(\frac{m_0}{2\pi T_0}\right)^{d/2}\exp\left(-\frac{m_0V^2}{2T_0}\right).
\end{equation}
The coefficients $a_0$, $b_0$ and $c_0$ are defined as
\begin{equation}
\label{4.5} a_0=-\frac{2}{d(d+2)}\frac{m_0}{n_0T_0^3}\int\;d{\bf v}\; {\bf S}_0({\bf
V})\cdot \boldsymbol{\mathcal{A}}_{0}({\bf V}),
\end{equation}
\begin{equation}
\label{4.6} b_0=-\frac{2}{d(d+2)}\frac{m_0}{n_0T_0^3}\int\;d{\bf v}\; {\bf S}_0({\bf
V})\cdot \boldsymbol{\mathcal{B}}_{0}({\bf V}),
\end{equation}
\begin{equation}
\label{4.7} c_{0}=-\frac{2}{d(d+2)}\frac{m_0}{n_0T_0^3}\int\;d{\bf v}\; {\bf S}_0({\bf
V})\cdot \boldsymbol{\mathcal{C}}_{0}({\bf V}).
\end{equation}
The transport coefficients $D_{0}$, $D$ and $D^T$ as well as the second Sonine
coefficients $a_0$, $b_0$ and  $c_0$ are determined by substitution of Eqs.\
(\ref{4.1})--(\ref{4.3}) into the integral equations (\ref{3.6})--(\ref{3.8}),
multiplication of these equations by $m_0{\bf V}$ and by ${\bf S}_0({\bf V})$, and
integration over velocity. The details are carried out in Appendices \ref{appA} and
\ref{appB} and only the final expressions will be presented here.

The second Sonine approximations $D_{0}[2]$, $D[2]$ and $D^T[2]$ can be written,
respectively, as
\begin{equation}
\label{4.8} D_{0}[2]=F(\alpha, \alpha_0, m_0/m, \sigma_0/\sigma,\phi) D_{0}[1],
\end{equation}
\begin{equation}
\label{4.9} D[2]=G(\alpha, \alpha_0, m_0/m, \sigma_0/\sigma, \phi) D[1],
\end{equation}
\begin{equation}
\label{4.10} D^T[2]=H(\alpha, \alpha_0, m_0/m, \sigma_0/\sigma, \phi) D^T[1],
\end{equation}
where $F$, $G$ and $H$ are nonlinear functions of the mass and size ratios, the
coefficients of restitution and the solid volume fraction. The explicit forms of $F$,
$G$ and $H$ are given by Eqs.\ (\ref{a.24}), (\ref{a.30}) and (\ref{a.22}),
respectively. In Eqs.\ (\ref{4.7})--(\ref{4.10}), $D_{0}[1]$, $D[1]$ and $D^T[1]$ refer
to the first Sonine approximations to $D_0$, $D$ and $D^T$, respectively. Their
explicit expressions were already determined in Ref.\ \cite{GHD07} for arbitrary
composition. In terms of the transport coefficients, the new calculations in the
present work are the functions $F$, $G$, and $H$. In the tracer limit ($x_0\to 0$), the
expressions of $D_{0}[1]$, $D[1]$ and $D^T[1]$ reduce, respectively, to
\begin{equation}
\label{4.12} D_{0}[1]=\frac{\rho
T}{m_0^2\nu}\frac{\gamma}{\nu_{1}^*-\frac{1}{2}\zeta^*},
\end{equation}
\begin{equation}
\label{4.13} D[1]=\frac{n_0 T}{m_0 \nu}\frac{Y_1^*}{\nu_{1}^*-\frac{1}{2}\zeta^*},
\end{equation}
\begin{equation}
\label{4.14} D^T[1]=\frac{n_0 T}{\rho \nu}\frac{X_1^*}{\nu_{1}^*-\zeta^*}.
\end{equation}
Here, $\nu=n\sigma^{d-1}\sqrt{2T/m}$ is an effective collision frequency,
$\gamma=T_0/T$ is the temperature ratio,
\begin{equation}
\label{4.16}
\zeta^{*}=\frac{\zeta^{(0)}}{\nu}=\frac{\sqrt{2}\pi^{(d-1)/2}}{d\Gamma(d/2)}\chi^{(0)}(1-\alpha^2)
\end{equation}
is the (reduced) cooling rate and the reduced quantities $X_1^*$, $Y_1^*$ and $\nu_1^*$
are given by Eqs.\ (\ref{a.21.1}), (\ref{a.28}) and (\ref{b.1}), respectively.

\begin{figure}
\includegraphics[width=0.5 \columnwidth,angle=0]{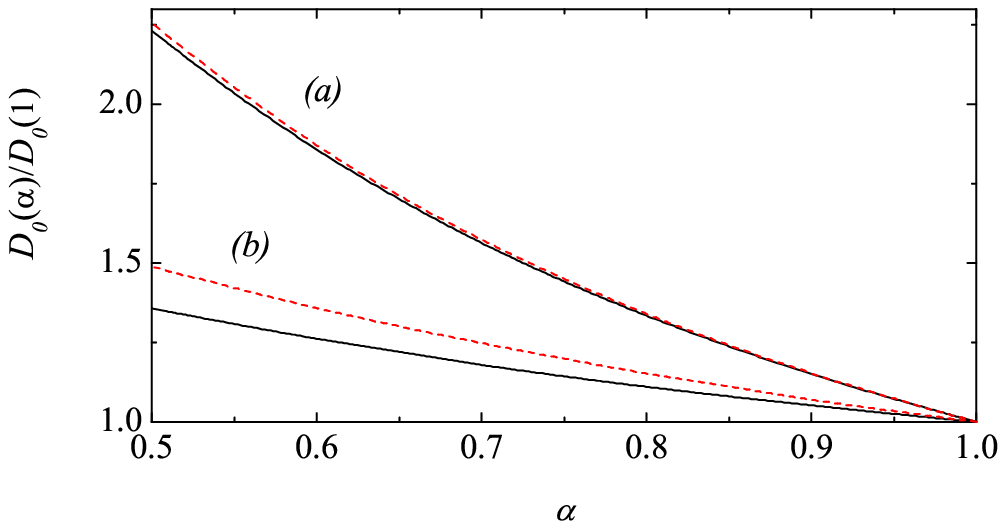}
\caption{(color online) Plot of the reduced kinetic diffusion coefficient
$D_{0}(\alpha)/D_{0}(1)$ as a function of the (common) coefficient of restitution
$\alpha=\alpha_0$ for the systems $(m_0/m=4,\sigma_0/\sigma=2)$ (a) and
$(m_0/m=0.5,\sigma_0/\sigma=0.8)$ (b) in the case of a three-dimensional gas with
$\phi=0.1$. The solid lines correspond to the second Sonine approximation while the
dashed lines refer to the first Sonine approximation. Here, $D_{0}(1)$ is the elastic
value of the kinetic diffusion coefficient consistently obtained in each approximation.
\label{fig1}}
\end{figure}
\begin{figure}
\includegraphics[width=0.5 \columnwidth,angle=0]{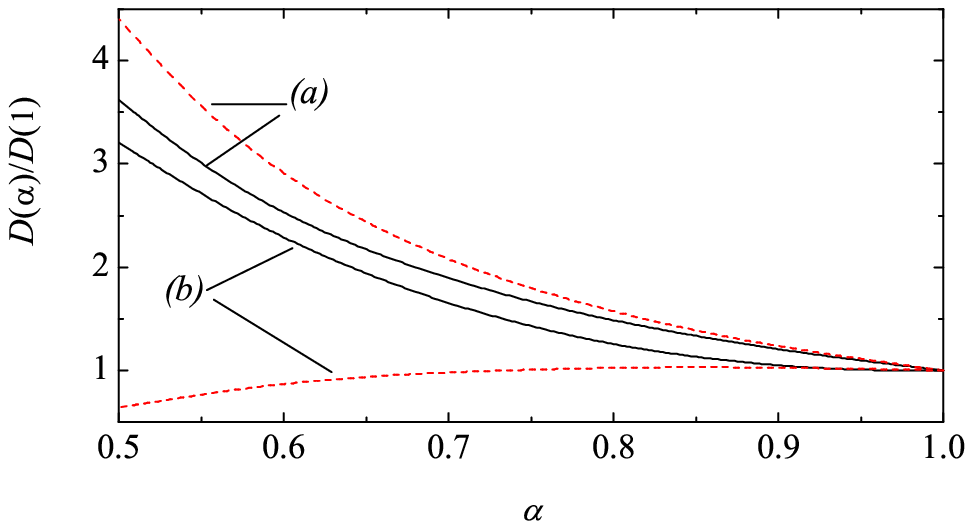}
\caption{(color online) Plot of the reduced mutual diffusion coefficient
$D(\alpha)/D(1)$ as a function of the (common) coefficient of restitution
$\alpha=\alpha_0$ for the systems $(m_0/m=4,\sigma_0/\sigma=2)$ (a) and
$(m_0/m=0.5,\sigma_0/\sigma=0.8)$ (b) in the case of a three-dimensional gas with
$\phi=0.1$. The solid lines correspond to the second Sonine approximation while the
dashed lines refer to the first Sonine approximation. Here, $D(1)$ is the elastic value
of the kinetic diffusion coefficient consistently obtained in each approximation.
\label{fig2}}
\end{figure}
\begin{figure}
\includegraphics[width=0.5 \columnwidth,angle=0]{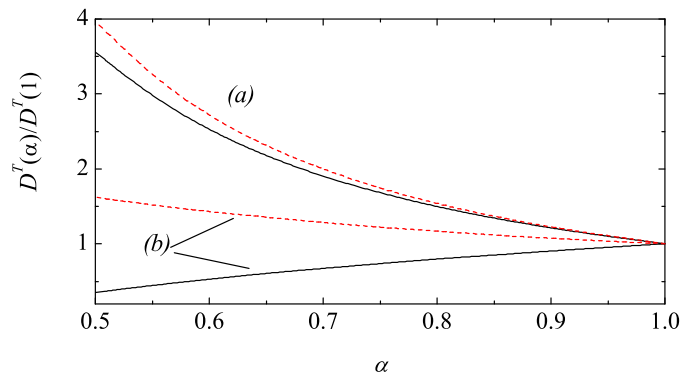}
\caption{(color online) Plot of the reduced thermal diffusion coefficient
$D^T(\alpha)/D^T(1)$ as a function of the (common) coefficient of restitution
$\alpha=\alpha_0$ for the systems $(m_0/m=4,\sigma_0/\sigma=2)$ (a) and
$(m_0/m=0.5,\sigma_0/\sigma=0.8)$ (b) in the case of a three-dimensional gas with
$\phi=0.1$. The solid lines correspond to the second Sonine approximation while the
dashed lines refer to the first Sonine approximation. Here, $D^T(1)$ is the elastic
value of the thermal diffusion coefficient consistently obtained in each approximation.
\label{fig3}}
\end{figure}

In general, the first and second Sonine approximations for the transport coefficients
of the mass flux have a complex dependence on the coefficients of restitution, the
solid fraction and the mass and size ratios. Thus, before analyzing this dependence, it
is instructive to consider some special limits. In the elastic limit
($\alpha=\alpha_0=1$) of a three-dimensional system, one recovers previous results for
a gas mixture of elastic hard spheres \cite{M54,MC84}. Moreover, in the case of
mechanically equivalent particles ($m_0=m$, $\sigma_0=\sigma$, $\alpha=\alpha_0$), as
expected, one gets $D^T[2]=0$, $D_{0}[2]=-(m/x_0m_0)D[2]$, and so
\begin{equation}
\label{4.17} {\bf j}_0^{(1)}=-\frac{nm_0^2}{\rho}D_{0}[2]\nabla x_0,
\end{equation}
where $x_0=n_0/n$ is the mole fraction of impurities. Moreover, in the case of a dilute
gas ($\phi=0$), the expression of the kinetic diffusion coefficient $D_{0}[2]$
coincides with the one previously derived by one of the authors \cite{GM04} by assuming
that the solvent is in the homogeneous cooling state. All these results confirm the
self-consistency of the results reported in this paper.

\begin{figure}
\begin{center}
\begin{tabular}{lr}
\resizebox{8cm}{!}{\includegraphics{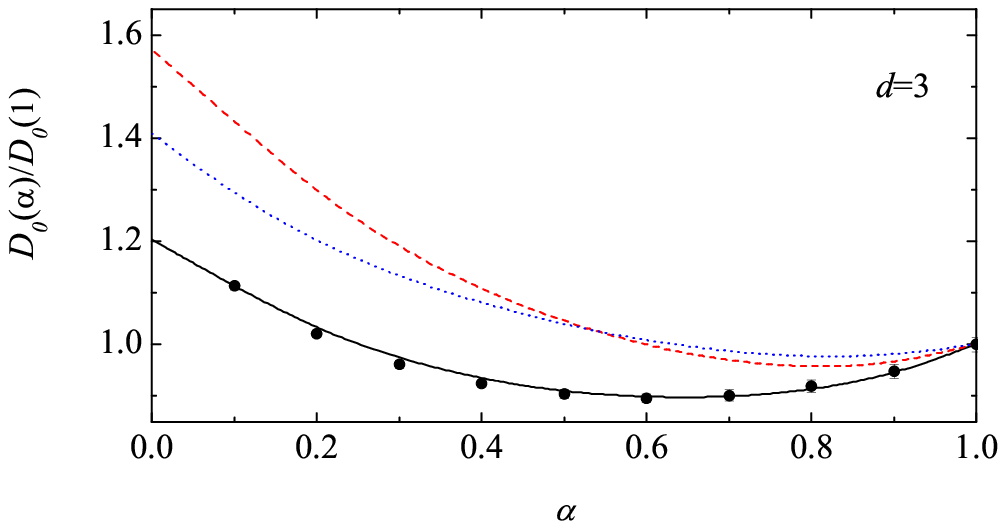}}&\resizebox{8cm}{!}{\includegraphics{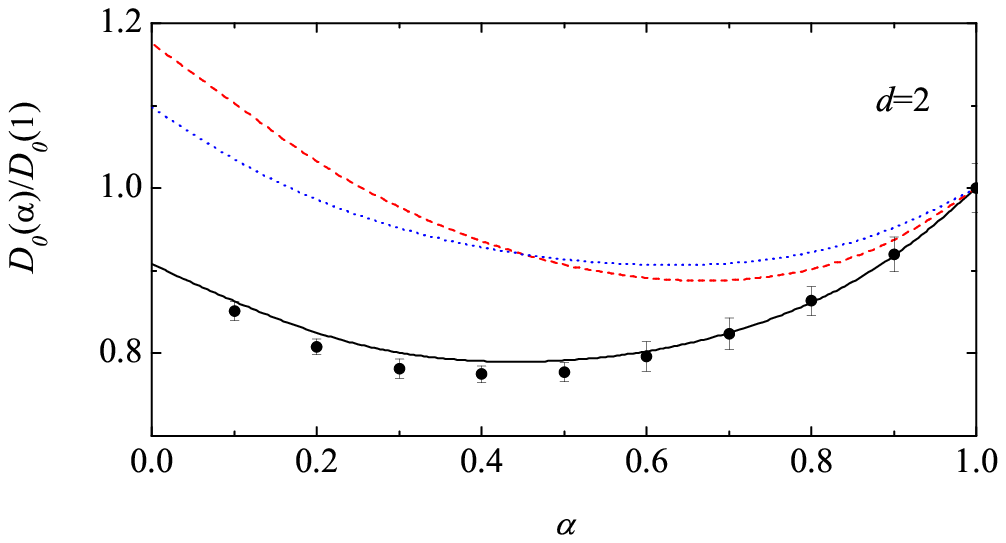}}
\end{tabular}
\end{center}
\caption{(color online) Plot of the reduced kinetic diffusion coefficient
$D_{0}(\alpha)/D_{0}(1)$ as a function of the (common) coefficient of restitution
$\alpha=\alpha_0$ for $m_0/m=1/8$, $\sigma_0/\sigma=1/2$ and $\phi=0$. The left panel
is for hard spheres $(d=3)$ while the right panel is for hard disks $(d=2)$. The solid
lines correspond to the second Sonine approximation, the dashed lines refer to the
first Sonine approximation and the dotted lines are the modified Sonine approximation.
The symbols are the results obtained from Monte Carlo simulations. Here, $D_{0}(1)$ is
the elastic value of the thermal diffusion coefficient consistently obtained in each
approximation. \label{fig9}}
\end{figure}

\begin{figure}
\begin{center}
\begin{tabular}{lr}
\resizebox{8cm}{!}{\includegraphics{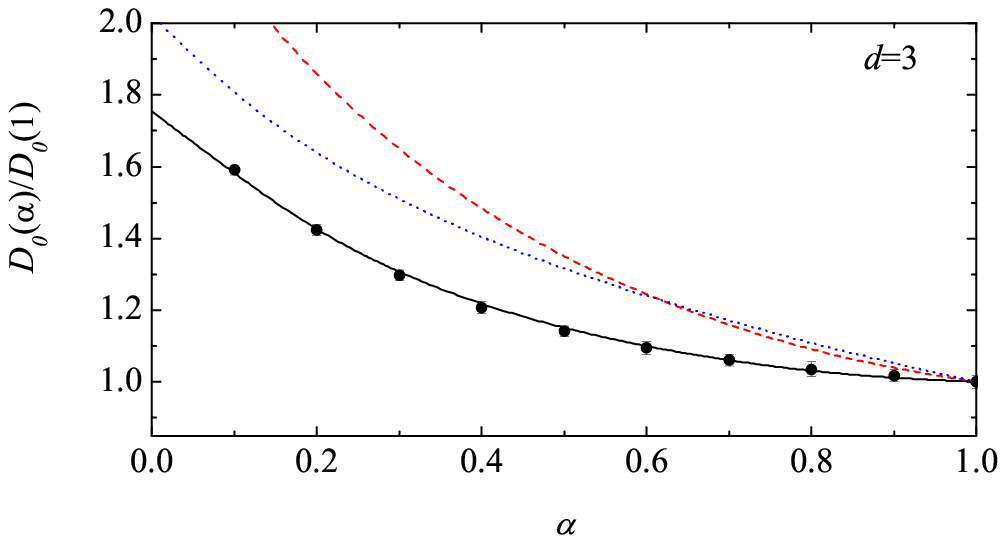}}&\resizebox{8cm}{!}{\includegraphics{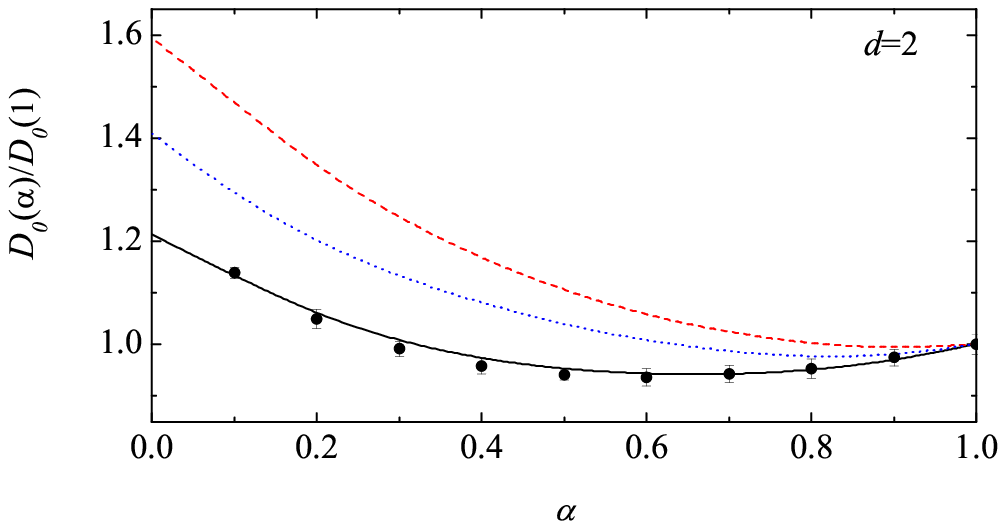}}
\end{tabular}
\end{center}
\caption{(color online) Plot of the reduced kinetic diffusion coefficient
$D_{0}(\alpha)/D_{0}(1)$ as a function of the (common) coefficient of restitution
$\alpha=\alpha_0$ for $m_0/m=1/5$, $\sigma_0/\sigma=1/2$ and $\phi=0.2$. The left panel
is for hard spheres $(d=3)$ while the right panel is for hard disks $(d=2)$. The solid
lines correspond to the second Sonine approximation, the dashed lines refer to the
first Sonine approximation and the dotted lines are the modified Sonine approximation.
The symbols are the results obtained from Monte Carlo simulations. Here, $D_{0}(1)$ is
the elastic value of the thermal diffusion coefficient consistently obtained in each
approximation. \label{fig10}}
\end{figure}
\begin{figure}
\begin{center}
\begin{tabular}{lr}
\resizebox{8cm}{!}{\includegraphics{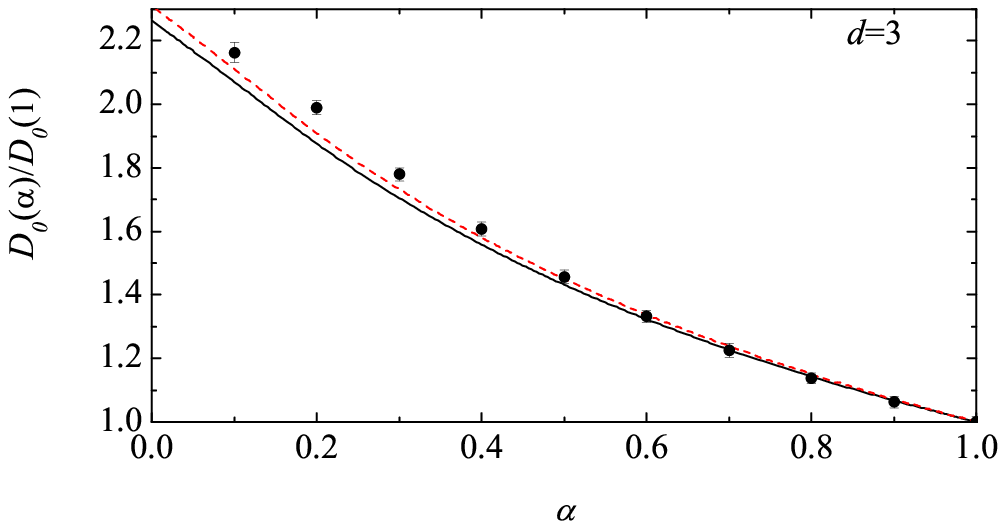}}&\resizebox{8cm}{!}{\includegraphics{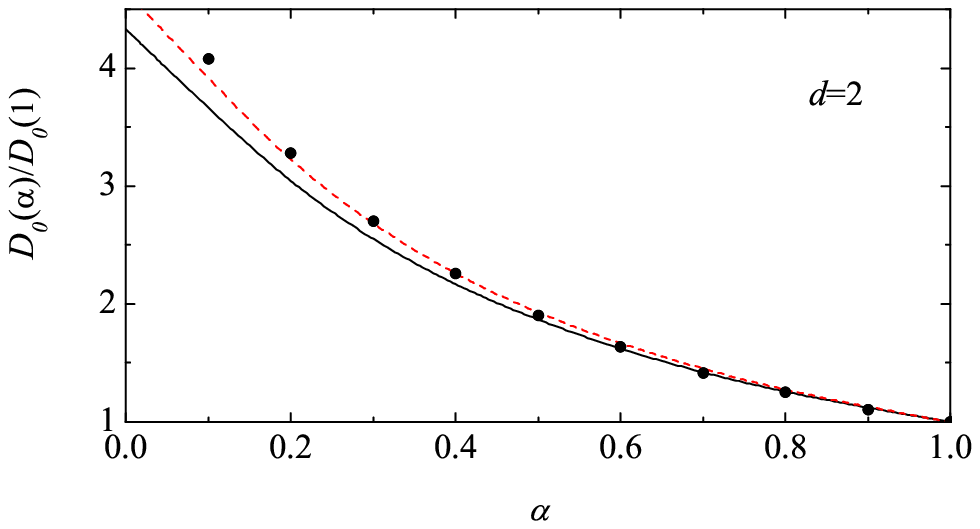}}
\end{tabular}
\end{center}
\caption{(color online) Plot of the reduced kinetic diffusion coefficient
$D_{0}(\alpha)/D_{0}(1)$ as a function of the (common) coefficient of restitution
$\alpha=\alpha_0$  for $m_0/m=2$, $\sigma_0/\sigma=2$ and $\phi=0.2$. The left panel is
for hard spheres $(d=3)$ while the right panel is for hard disks $(d=2)$. The solid
lines correspond to the second Sonine approximation and the dashed lines refer to the
first Sonine approximation. The symbols are the results obtained from Monte Carlo
simulations. Here, $D_{0}(1)$ is the elastic value of the thermal diffusion coefficient
consistently obtained in each approximation. \label{fig11}}
\end{figure}

\section{Some illustrative examples and comparison with Monte Carlo simulations}
\label{sec5}

The expressions (\ref{4.8})--(\ref{4.10}) for the (reduced)  transport coefficients
$D_0[2]/D_0[1]$, $D[2]/D[1]$ and $D^T[2]/D^T[1]$ depend on many parameters: $\{m_0/m,
\sigma_0/\sigma, \alpha,\alpha_0, \phi\}$, or equivalently \cite{SD06} $\{m_0^*/m,
\sigma_0/\sigma, \alpha,\phi \}$, where $m_0^*$ is defined by Eq.\ (\ref{2.15.0}). This
complexity exists in the elastic limit as well \cite{LCK83}, except for the dependence
on the coefficients of restitution. Thus, to show more clearly the influence of
dissipation on the transport coefficients we normalize them with respect to their
values in the elastic limit. Also, for simplicity, we take the simplest case of common
coefficient of restitution $\alpha=\alpha_0$ so that, the parameter space has four
independent quantities: $\{m_0/m, \sigma_0/\sigma, \alpha, \phi\}$.

In order to get the explicit dependence of the transport coefficients on the above four
parameters one has to give the forms of the pair correlation functions $\chi^{(0)}$ and
$\chi_0^{(0)}$. In the three-dimensional case ($d=3$), a good approximation for
$\chi^{(0)}$ is provided by the Carnahan-Starling form \cite{CS69}
\begin{equation}
\label{n.1} \chi^{(0)}=\frac{1-\frac{1}{2}\phi}{(1-\phi)^3},
\end{equation}
while the intruder-gas pair correlation function is given by \cite{B70}
\begin{equation}
\label{n.2}
\chi_0^{(0)}=\frac{1}{1-\phi}+3\frac{\omega}{1+\omega}\frac{\phi}{(1-\phi)^2}+2
\frac{\omega^2}{(1+\omega)^2}\frac{\phi^2}{(1-\phi)^3},
\end{equation}
where we recall that $\omega=\sigma_0/\sigma$ is the diameter ratio. The expression for
the chemical potential of the intruder consistent with the approximation (\ref{n.2}) is
\cite{RG73}
\begin{eqnarray}
\label{n.3} \frac{\mu_0}{T}&=&C_3+\ln n_0-\ln (1-\phi)+3\omega \frac{\phi}{1-\phi}+
3\omega^2\left[\ln (1-\phi)+\frac{\phi(2-\phi)}{(1-\phi)^2}\right]\nonumber\\
& & -\omega^3\left[2\ln (1-\phi)+\frac{\phi(1-6\phi+3\phi^2)}{(1-\phi)^3}\right],
\end{eqnarray}
where $C_3$ is a constant. For a two-dimensional gas ($d=2$), $\chi^{(0)}$ and
$\chi_0^{(0)}$ are approximately given by \cite{JM87}
\begin{equation}
\label{n.4} \chi^{(0)}=\frac{1-\frac{7}{16}\phi}{(1-\phi)^2},
\end{equation}
\begin{equation}
\label{n.5}
\chi_0^{(0)}=\frac{1}{1-\phi}+\frac{9}{8}\frac{\omega}{1+\omega}\frac{\phi}{(1-\phi)^2}.
\end{equation}
The chemical potential is now given by \cite{S08}
\begin{equation}
\label{n.6}
\frac{\mu_0}{T}=C_2+\ln n_0-\ln (1-\phi)+\frac{1}{4}\omega \left[\frac{9\phi}{1-\phi}+\ln(1-\phi)\right]+
\frac{1}{8}\omega^2\left[
\frac{\phi(7+2\phi)}{(1-\phi)^2}-\ln(1-\phi)\right],
\end{equation}
where $C_2$ is a constant.

In Figs.\ \ref{fig1}--\ref{fig3}, we plot the transport coefficients for inelastic hard
spheres ($d=3$) as functions of the coefficient of restitution for two different
systems. Each transport coefficient has been reduced with respect to its elastic value
consistently obtained in each approximation. The dashed lines refer to the first Sonine
approximation while the solid lines correspond to the second Sonine approximation. We
observe that in general the first Sonine polynomial approximation quantitatively
differs from the second Sonine approach as the dissipation increases for sufficiently
small values of the mass ratio $m_0/m$ and/or the size ratio $\sigma_0/\sigma$. For
these cases, the first Sonine approximation is not sufficient to capture the influence
of dissipation on mass transport. However, the predictions of the first Sonine
correction improve significantly as $m_0/m$ and/or $\sigma_0/\sigma$ increases so that
the former accurately describes the mass transport in this range of values of the mass
and size ratios, even for strong inelasticity. These findings on the convergence of the
Sonine polynomial expansion are quite similar to those obtained for elastic systems
\cite{MC84} and for granular gases at low-density \cite{GM04}.

To check the reliability of the first and second Sonine approximations, we have
performed Monte Carlo simulations of the Enskog equation for the dense granular gas
with the tracer particles. We have extracted from these simulations the kinetic
diffusion coefficient $D_0$ of impurities in a granular dense gas in the homogeneous
cooling state (HCS). This coefficient can be obtained from the mean square displacement
of the intruder particle after a time interval $t$ as \cite{GM04,McL89}
\begin{equation}
\label{n.4.1} \frac{\partial}{\partial t}\langle |{\bf r}(t)-{\bf
r}(0)|^2 \rangle =\frac{2dD_0}{n},
\end{equation}
where $|{\bf r}(t)-{\bf r}(0)|$ is the distance travelled by the intruder from $t=0$
until time $t$. Equation (\ref{n.4.1}) is the Einstein form of the diffusion
coefficient. This relation can be used also in Monte Carlo simulations of granular
gases to measure the diffusion coefficient (see for example a previous work on transport of impurities in a dilute granular gas in Ref.\ \cite{GM04}). In an
unbounded system like ours, the DSMC method has two steps that are repeated in each
time iteration: the first step takes care of the particles drift and the second step
accounts for the collisions among particles. The extension of the DSMC method to study
the diffusion of impurities in a dense granular gas in the HCS requires the changes
$J[f,f]\to \chi J[f,f]$ and $J_0[f_0,f]\to \chi_0 J_0[f_0,f]$. For the DSMC method to
work appropriately, the time step needs to be small in comparison with the microscopic
time scale of the problem (which is set by the collision frequency $\nu$) and we also
need a sufficiently high number of simulated particles \cite{B94}. Thus, we have used
in the simulations of this work a time step $\delta t=2.5\times10^{-4}\nu^{-1}$ and
$N=2\times 10^6$ simulated particles for each species \cite{DSMCnote}. To our knowledge
and since we are interested in the complete range of values of $\alpha$, we present the
first DSMC data on dense granular gases for coefficients of restitution as low as
$\alpha=0.1$. More details on the application of the DSMC method \cite{B94} to this
diffusion problem can be found in Ref.\ \cite {GM04}.

If a hydrodynamic description (or normal solution in the context
of the CE method) applies, then the diffusion coefficient $D_0(t)$
depends on time only through its dependence on the temperature
$T(t)$. In this case, after a transient regime, the reduced
diffusion coefficient $D_0(\alpha)/D_0(1)$
achieves a time-independent value \cite{GM04}. Here, we compare the steady
state values of $D_0(\alpha)/D_0(1)$ obtained from Monte Carlo simulations with
the theoretical predictions given by the first and second Sonine
approximations.

Let us consider first the dilute gas limit ($\phi=0$). Figure \ref{fig9} shows the
reduced diffusion coefficient $D_0(\alpha)/D_0(1)$ for $m_0/m=1/8$ and
$\sigma_0/\sigma=1/2$ for disks ($d=2$) and spheres ($d=3$). According to the theory
results obtained in the previous figures, one expects that in the Lorentz gas limit
(small values of the mass and size ratios) both Sonine approximations differ
significantly for strong inelasticity (this means obviously that it is not possible
that both of them simultaneously show good agreement with the exact solution of the
problem). For the sake of completeness, we have also included the results recently
derived  for binary mixtures \cite{GVM09} from a new method based on a modified version
of the first Sonine approximation which replaces the Maxwell-Boltzmann distribution
weight function (used in the standard Sonine approximations) by the homogeneous cooling
state distribution \cite{GSM07,GVM09}. This new method partially eliminates the
observed disagreement (for strong dissipation) between computer simulations \cite{BR05}
and theoretical results for the heat flux transport coefficients \cite{GVM09}. The
explicit expression for the coefficient $D_0$ obtained from the modified Sonine method
is displayed in Appendix \ref{appD}. The simulation data corresponding to $d=3$ for
$\alpha \geq 0.5$ were reported in Ref.\ \cite{GM04} while those corresponding to $d=2$
and $d=3$ for $\alpha \leq 0.5$ have been obtained in this work. It is quite apparent
that while the second Sonine approximation agrees very well with simulation data, the
standard and modified first Sonine approximations fail for strong dissipation. This
agreement is specially significant in the case of hard spheres. Thus, as expected, the
Sonine polynomial expansion exhibits a poor convergence for sufficiently small values
of the mass and size ratios. To assess the influence of density on these trends, the
ratio $D_0(\alpha)/D_0(1)$ is plotted in Fig.\ \ref{fig10} for $m_0/m=1/5$,
$\sigma_0/\sigma=1/2$ in the case of a moderately dense gas ($\phi=0.2$). As before,
both first Sonine approximations underestimate the diffusion coefficient (the
discrepancy being more important in the case of the standard than the modified first
Sonine approximation) while the predictions of the second Sonine approach show an
excellent agreement with simulation data in the whole range of values of the
coefficient of restitution. All these results clearly confirm the accuracy of the
second Sonine approximation, even for low values of $\alpha$ and small values of the
mass and size ratios. Figure \ref{fig11} shows that in the opposite limit  of large
values of the mass and size ratios (Rayleigh gas limit), the first and second Sonine
approximations are practically indistinguishable for moderately large inelasticity (say
for instance, $\alpha \geq 0.5$) and both approaches provide a general good agreement
with Monte Carlo simulations. However, at very high inelasticity, the second Sonine
approximation very slightly underestimates the diffusion coefficient compared to the
DSMC data and the first Sonine approximation. We have performed more series of
simulations (not shown here) with different values of the ratios $m_0/m$ and
$\sigma_0/\sigma$ confirming similar trends as those in the figures shown in this
Section for both cases $m_0/m<1$ and/or $\sigma_0/\sigma<1$ and $m_0/m>1$ and/or
$\sigma_0/\sigma>1$ (the data are available to the reader upon request to the authors).

\section{Segregation by thermal diffusion}
\label{sec6}

As an application of the previous results, this Section is devoted to the study of
segregation, driven by both gravity and temperature gradients, of an intruder in a
granular dense gas. Segregation and mixing of dissimilar grains is one of the most
interesting problems in granular mixtures not only from a fundamental point of view but
also from a more practical point of view. This problem has spawn a number of important
experimental, computational and theoretical works in the field of granular media
\cite{K04}. Although several mechanisms have been proposed in the literature, the
problem is not completely understood yet. Among the different mechanisms, thermal
diffusion becomes the most relevant if the system resembles the conditions of a
granular gas. In this case, kinetic theory tools have proven to be quite useful to
analyze the motion of the intruder. A short previous analysis on this problem, but when
the system is heated by a stochastic-driving force, has been reported in Ref.\
\cite{G08}.

\begin{figure}
\includegraphics[width=0.5 \columnwidth,angle=0]{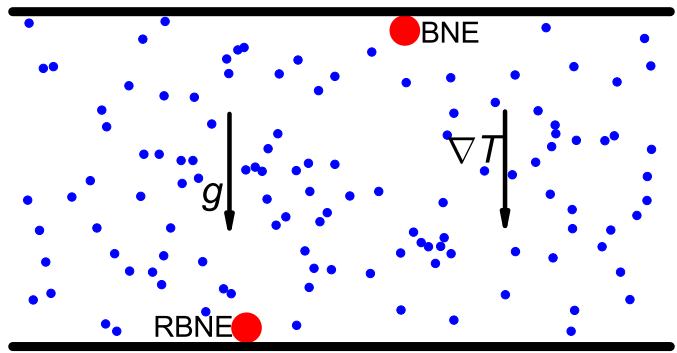}
\caption{(color online) Sketch of the segregation problem analyzed in Sec.\ \ref{sec6}. The small circles represent
the particles of the dense gas while the large circles are the intruders. The BNE
(RBNE) effect corresponds to the situation in which the intruder rises (falls) to the
top (bottom) plate. \label{fig4}}
\end{figure}

Thermal diffusion is caused by the relative motion of the components of a mixture due
to the presence of a temperature gradient. As a result of this motion, a steady state
is finally reached in which the separating effect arising from thermal diffusion is
balanced by the remixing effect of ordinary diffusion \cite{KCL86}. Under these
conditions, the thermal diffusion factor $\Lambda$ characterizes the amount of
segregation parallel to the temperature gradient. Our goal here is to determine
$\Lambda$ in a non-convecting (${\bf u}={\bf 0}$) steady state with gradients only
in the vertical direction ($z$ axis) for simplicity. In this case, $\Lambda$ is
defined as
\begin{equation}
\label{5.1}
-\Lambda\partial_z \ln T=\partial_z\ln \left(\frac{n_0}{n}\right).
\end{equation}
If we assume that gravity and thermal gradient point in parallel directions (i.e., the
bottom plate is hotter than the top plate), then the intruder rises with respect to the
fluid particles if $\Lambda>0$ while the intruder falls with respect to the fluid
particles if $\Lambda<0$. If the impurity is heavier than the gas particles, the former
situation is referred to as the Brazil-nut effect (BNE) while the latter is called the
reverse Brazil-nut effect (RBNE). A sketch of the geometry of the segregation problem
studied here is given in Fig.\ \ref{fig4}. The key point here is that logically the
segregation condition (\ref{5.1}) will depend on the mass flux transport coefficients
of the intruder. We remind the reader of the features, some of them described in the
Introduction, that the transport coefficients we use here are able to capture (compared to alternative
coefficients in theoretical works on segregation by other authors \cite{JY02,TAH03}). These
features are a consequence of the consistent development of the Chapman-Enskog theory
for dense granular mixtures \cite{GDH07}. Due to this, and in addition to the gain of accuracy with the use of the second order term in the Sonine expansion, we expect that  our expressions \eqref{4.12}-\eqref{4.14} will be useful in practical problems and applications of granular segregation \cite{MATHnote}. We determine below the explicit expression of this segregation criterion.

As said before, let us consider a steady base state with no flow (${\bf u}={\bf 0}$)
and with a temperature gradient parallel to the direction of gravity (in this case, the
$z$ direction). According to Eq.\ (\ref{2.16}), the mass flux ${\bf j}_0$ vanishes in
this state (because ${\bf u}={\bf 0}$) and there are no contributions to the pressure
tensor except for those coming from the hydrostatic pressure term, i.e.,
$P_{ij}=p\delta_{ij}$, where $p$ is given by Eq.\ (\ref{3.12.1}). As a consequence, the
momentum balance equation (\ref{2.8}) becomes
\begin{equation}
\label{5.2} \frac{\partial p}{\partial z}=\frac{\partial p}{\partial
T}\partial_zT+\frac{\partial p}{\partial n}\partial_z n=-\rho g.
\end{equation}
Finally, the constitutive equation for the mass flux is given by Eq.\ (\ref{1.1}) with
$\nabla \to \partial_z$. Using the fact that $j_{0,z}=0$ and taking into account Eqs.\
(\ref{1.1}) and (\ref{5.1}), the factor $\Lambda$ can be written as
\begin{equation}
\label{5.3}
\Lambda=\frac{\beta D^{T*}-(p^*+g^*)(D_0^*+D^*)}{\beta D_0^*},
\end{equation}
where we have introduced the reduced transport coefficients $D^{T*}=(\rho\nu/n_0T)D^T$,
$D_0^*=(m_0^2\nu/\rho T)D_0$, and $D^*=(m_0\nu/n_0T)D$. Moreover,
$p^*=p/nT=1+2^{d-2}(1+\alpha)\chi^{(0)} \phi$,
\begin{equation}
\label{5.3.1}
\beta=p^*+\phi \partial_\phi p^*=1+2^{d-2}(1+\alpha)\chi^{(0)} \phi \left[1+\phi\frac{\partial}
{\partial \phi}\ln (\phi \chi^{(0)})\right]
\end{equation}
and
\begin{equation}
\label{5.4}
g^*=\frac{\rho g}{n\left(\frac{\partial T}{\partial z}\right)}<0
\end{equation}
is a dimensionless parameter measuring the gravity relative to the thermal gradient.
This parameter measures the competition between these two mechanisms ($g$ and
$\partial_z T$) on segregation.

\begin{figure}
\includegraphics[width=0.4 \columnwidth,angle=0]{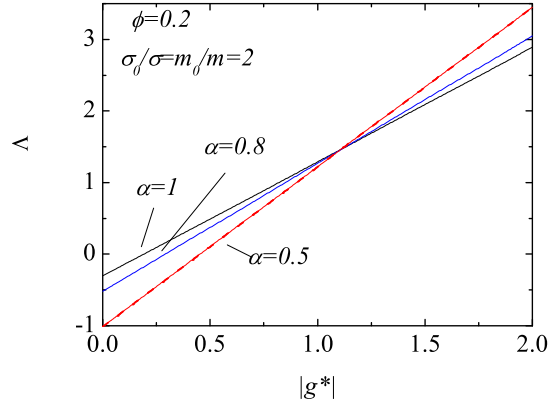}
\caption{(color online) Plot of the thermal diffusion factor $\Lambda$ versus the
(reduced) gravity $|g^*|$ for $\phi=0.2$ and three values of the (common) coefficient
of restitution $\alpha$: $\alpha=1$, 0.8 and 0.5. \label{fig6}}
\end{figure}

As expected, the dependence of $\Lambda$ on the parameter space of the problem is quite
intricate. Regarding the dependence of thermal diffusion on gravity, we observe that
for given values of the mass and size ratios, the coefficient of restitution and
density, it is possible to switch between RBNE ($\Lambda<0$) and BNE ($\Lambda>0$) by
changing the value of gravity relative to the thermal gradient. This is a new
interesting effect not captured in previous works on segregation
\cite{JY02,TAH03,ATH06}. As an illustration of this effect, Fig.\ \ref{fig6} presents
plots of the second Sonine approximation for $\Lambda$ as a function of the (reduced)
gravity for a three-dimensional system with $\phi=0.2$, $\sigma_0/\sigma=m_0/m=2$ and
three different values of the (common) coefficient of restitution $\alpha$.  It is
apparent that, for the case analyzed here, the RBNE is dominant at small values of
$|g^*|$ while the opposite happens as the dimensionless gravity increases.

\begin{figure}
\includegraphics[width=0.4 \columnwidth,angle=0]{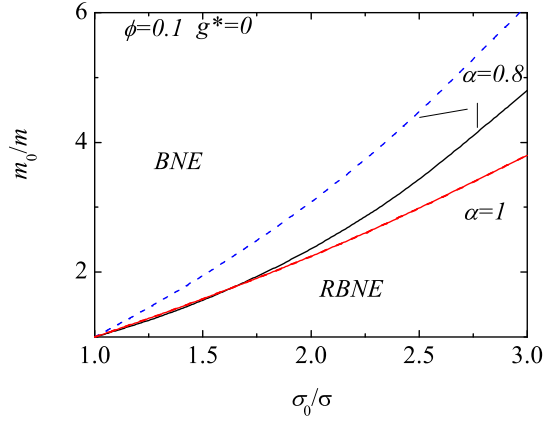}
\caption{(color online) Phase diagram for BNE/RBNE for $\phi=0.1$ in the absence of
gravity and for two values of the (common) coefficient of restitution $\alpha$. Points
above the curve correspond to $\Lambda >0$ (BNE) while points below the curve
correspond to $\Lambda <0$ (RBNE). The dashed line is the result obtained from the
first Sonine approximation for $\alpha=0.8$. \label{fig7}}
\end{figure}
\begin{figure}
\includegraphics[width=0.4 \columnwidth,angle=0]{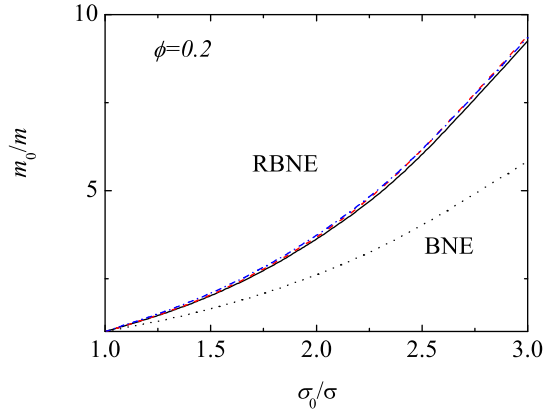}
\caption{(color online) Phase diagram for BNE/RBNE for $\phi=0.2$ in the absence of
thermal gradient ($|g^*|\to \infty$) for three values of $\alpha$: $\alpha=1$, $0.8$
(dashed line) and $0.5$ (dashed-dotted line). The dotted line refers to the results
obtained by Jenkins and Yoon \cite{JY02} for an elastic system. \label{fig8}}
\end{figure}

The condition $\Lambda=0$ provides the segregation criterion for the transition BNE
$\Leftrightarrow$ RBNE. Since $\beta$ and $D_0^*$ are positive, then, according to
(\ref{5.3}), $\text{sgn} (\Lambda)=\text{sgn}(\beta D^{T*}-(p^*+g^*)(D_{0}^*+D^*))$. As
a consequence, the segregation criterion is
\begin{equation}
\label{5.5}
\beta D^{T*}= (p^*+g^*)(D_{0}^*+D^*).
\end{equation}
As expected, when $m_0=m$, $\sigma_0=\sigma$ and $\alpha=\alpha_0$, the system
(intruder plus gas) is monodisperse and the two species do not segregate. This is
consistent with Eq.\ (\ref{5.5}) since in this limit case $D^{T*}=D_0^*+D^*=0$ so that
$\Lambda=0$ for any value of $\alpha$ and $\phi$. On the other hand, in the case of a dilute gas ($\phi=0$), one has
$\beta=p^*=1$ and the condition (\ref{5.5}) when $g^*=0$ in the first Sonine approximation becomes
\begin{equation}
\label{5.6}
\frac{m_0}{m}=\frac{T_0}{T}.
\end{equation}
This segregation condition agrees with some recent results derived from the Boltzmann
equation \cite{BRM05,G06}. It must be remarked that, due to the lack of energy
equipartition, the condition $m_0/m=T_0/T$ is rather complicated since it involves all
the parameters of the system.

We consider now dense systems. For the sake of concreteness, we assume that the
intruder is larger than the gas particles ($\sigma_0>\sigma$). Figure \ref{fig7} shows
a phase-diagram in the $(m_0/m, \sigma_0/\sigma)$-plane for a moderate dense gas
($\phi=0.1$) in the absence of gravity and two values of the coefficient of
restitution. Also, for comparison the corresponding phase-diagram obtained from the
first Sonine approximation is plotted for the case $\alpha=0.8$. We observe that, in
the absence of gravity, the main effect of dissipation is to reduce the size of the
BNE. This conclusion qualitatively agrees with the results derived in the driven gas
case \cite{G08}. However, at a quantitative level, the influence of dissipation
observed here is less important than the one obtained in the heated case. Moreover,
although the first Sonine approximation reproduces qualitatively the trends of the
phase-diagram, the former overestimates the predictions of the second Sonine
approximation, especially when increasing the mass and size ratios.

In some previous theoretical studies \cite{JY02,TAH03,ATH06}, it has been assumed that
the global temperature of the bed does not vary with height so that the effect of
$\partial_z T$ on segregation is neglected. In this limit ($|g^*|\to \infty$), the
condition (\ref{5.5}) reduces to $D_0^*+D^*=0$. The form of the phase diagram in the
limit $|g^*|\to \infty$ is shown in Fig.\ \ref{fig8} for $\phi=0.2$ and three values of
the coefficient of restitution. The result derived by Jenkins and Yoon \cite{JY02} in
the elastic case from a simple kinetic theory (namely, just a particular case from the
perspective of the theory presented in this work) has also been included for
comparison. In contrast to the case $g^*=0$, we observe that the RBNE regime appears
essentially now for both large mass ratio and/or small size ratio. With respect to the
influence of inelasticity, Fig.\ \ref{fig8} shows that the phase diagram is practically
independent of the value of the coefficient of restitution since the three curves
collapse in a common curve. We also observe that our results differ from those obtained
by Jenkins and Yoon \cite{JY02}, especially for large size ratios.

\section{Conclusions}
\label{sec7}

In this paper we have analyzed the mass transport of impurities in a moderately dense
granular gas described by the inelastic Enskog kinetic equation. This is perhaps the
simplest example of transport in a \emph{multicomponent} granular gas since the tracer
particles (impurities) are enslaved to the granular gas (solvent) and there are fewer
parameters. Nevertheless, it involves, as we explained, many situations of practical
interest in the study of granular gases. In the tracer limit, once the state of the
solvent is well characterized, the mass flux ${\bf j}_0$ associated with impurities is
the relevant flux of the problem. To first order in the spatial gradients, ${\bf j}_0$
is given by Eq.\ (\ref{1.1}) where $D_0$ is the kinetic diffusion coefficient, $D$ is
the mutual diffusion coefficient and $D^T$ is the thermal diffusion coefficient.

The main goal of this paper has been to determine these three transport coefficients as
functions of the temperature, the density and the different mechanical parameters of
the system, namely, the masses and particle diameters and the (constant) coefficients
of restitution for the impurity-gas and gas-gas collisions. Like for elastic collisions
\cite{LCK83}, the coefficients $D_0$, $D$ and $D^T$ are given in terms of the solutions
of a set of coupled linear integral equations \cite{GDH07}. A practical evaluation of
the above diffusion coefficients is possible by using a Sonine polynomial expansion and
approximate results are not limited to weak inelasticity. Here, $D_0$, $D$ and $D^T$
have been determined in the first (one polynomial) and second (two polynomials) Sonine
approximation and progress was possible here thanks to previous results obtained by
using the (standard) first Sonine approximation \cite{GHD07} for the full Navier-Stokes
transport coefficients of polydisperse dense mixtures. The present study complements
and extends previous works on diffusion in granular dilute \cite{GM04} and dense
\cite{LBD02} gases and provides explicit expressions for $D_0$, $D$ and $D^T$ beyond
the first Sonine approximation \cite{GHD07}.

Comparison of the theoretical results derived for $D_0$, $D$ and $D^T$ between the
first and second Sonine approximations shows significant discrepancies between both
approaches for values of the mass ratio $m_0/m$ and/or the size ratio $\sigma_0/\sigma$
smaller than 1 while the quality of the first Sonine correction improves with
increasing values of $m_0/m$ and $\sigma_0/\sigma$. These trends are quite similar to
those previously found for tracer diffusion in an ordinary (elastic) dense gas
\cite{MC84} and in a granular dilute gas \cite{GM04}. Moreover, to check to reliability
of the different theoretical approaches, a comparison with Monte Carlo simulations of
the Enskog equation for the coefficient $D_0$ has been carried out for disks ($d=2$)
and spheres ($d=3$). The comparison with simulation data shows the superiority of the
second Sonine approximation over other approaches (the standard first Sonine
approximation and a modified version of the first Sonine correction recently proposed
\cite{GSM07,GVM09}), since the agreement of $D_0[2]$ with numerical results is
excellent, even for strong dissipation (see for instance Figs.\ \ref{fig9} and
\ref{fig10}) and very small values of the mass and size ratios.

With respect to the segregation problem, which has of growing interest in the research
community in the field, we have shown that the explicit knowledge of the three
diffusion coefficients allows one to compute the thermal diffusion factor $\Lambda$.
This quantity provides a convenient measure of the separation or segregation generated
by a temperature gradient in a multicomponent system. According to the symmetry of the
problem (sketched in Fig.\ \ref{fig4}), when $\Lambda>0$ the intruder tends to climb to
the top of the sample against gravity (Brazil-nut effect, BNE) while if $\Lambda<0$ the
intruder tends to move at the bottom of the system (reverse Brazil-nut effect, RBNE).
The understanding of the transition BNE/RBNE is of central interest in the field of
granular  matter mainly due to its   practical/industrial importance. The analysis
carried out here provides an extension of a previous analysis \cite{G08} for a heated
dense gas in the first Sonine approximation. Our results show that the influence of
dissipation on the phase diagram BNE/RBNE is much more significant in the absence of
gravity ($|g^*|=0$) than in the opposite limit ($|g^*|=\infty$). In fact, as Fig.\
\ref{fig8} shows, when the segregation of the intruder is essentially driven by
gravity, the inelasticity of collisions has not discernible influence on the form of
the phase diagram BNE/RBNE. From the discussion in previous Sections, we expect the
segregation criteria presented here to be more accurate in comparison with the criteria
derived in other works \cite{JY02,TAH03}. Future research work using a DSMC code
adapted to the problem of segregation, molecular dynamics (MD) simulations, and
eventually, experiments, will help us to test our theory (and previous alternative
theories) in real problems. We are currently working on DSMC and MD simulations.

An important issue is the usefulness of the expressions for the NS transport coefficients derived here. As already said in a previous work \cite{GVM09}, the NS hydrodynamic equations themselves may or may not be limited with respect to inelasticity, depending on the particular granular flow considered. In particular, tracer diffusion in the HCS at very low values of the coefficient of restitution is only possible for very small systems due to the spontaneous formation of velocity vortices and density clusters. Moreover, in most of problems of practical interest (such as steady states for a granular gas heated or sheared from the boundaries), the strength of spatial gradients is set by inelasticity so that the NS description only holds in the quasielastic limit \cite{SGD04}. Nevertheless, in spite of the above cautions, the NS equations are still appropriate for a wide class of flows. Some of them correspond to the stability analysis of small perturbations of the HCS  \cite{BMC99}, supersonic flows past a wedge \cite{RBSS02} and hydrodynamic profiles of systems vibrated vertically \cite{YHCMW02} where comparisons between theory and experiments have shown both qualitative and quantitative agreement for moderate values of dissipation (say for instance, $\alpha \gtrsim 0.8$). Consequently, the NS equations with the transport coefficients derived here can be considered still as an useful theory for a wide class of rapid granular flows, although more limited than for ordinary gases.

One of the main limitations of the present study is its restriction to the tracer or
intruder limit. This precludes the possibility of analyzing the influence of
composition on the mass transport. The extension of the results derived here to finite
mole fractions is an interesting open problem. Moreover, it would also be interesting
to evaluate the expressions for the remaining transport coefficients of the mixture
(shear viscosity, bulk viscosity, thermal conductivity, $\ldots$) for a variety of mass
and diameter ratios. This evaluation would allow one to assess the quality of the
approximate Sonine method for solving the integral equations for the transport
coefficients through a comparison with computer simulations. Previous results obtained
for the shear viscosity coefficient \cite{GM03} have shown a good agreement. Moreover,
the knowledge of the full NS transport coefficients for a dense granular binary mixture
allows us to determine the dispersion relations for the hydrodynamic equations
linearized about the homogeneous cooling state. Some previous results \cite{GMD06}
based on the Boltzmann kinetic equation  have shown that the resulting equations
exhibit a long wavelength instability for three of the modes. The objective now is to
extend to higher densities this previous linear stability analysis for a dilute gas
\cite{GMD06} and compare the theoretical predictions with MD simulations for the
homogenous cooling state. We plan to carry out such studies in the near future.

\acknowledgments

We are grateful to Andr\'es Santos for useful comments on an early version of this paper.
This research has been supported by the Ministerio de Educaci´on y Ciencia (Spain)
through Programa Juan de la Cierva (F.V.R.) and Grant No. FIS2007--60977. Partial
support from the Junta de Extremadura through Grant No. GRU07046 is also acknowledged.

\appendix
\section{First order velocity distribution function of the gas}
\label{appC}

The first order velocity distribution function $f^{(1)}$ of the gas particles has the
form \cite{GD99a,L05}
\begin{equation}
\label{c.1} f^{(1)}=\boldsymbol{\mathcal{A}}\cdot \nabla T+\boldsymbol{\mathcal{C}}\cdot \nabla n+
\boldsymbol{\mathcal{D}}:\nabla {\bf u}+E \nabla\cdot {\bf u}.
\end{equation}
Only the coefficients $\boldsymbol{\mathcal{A}}$ and $\boldsymbol{\mathcal{C}}$ are
involved in the evaluation of the mass transport ${\bf j}_0^{(1)}$ of the intruder.
These quantities verify the linear integral equations \cite{GD99,L05}
\begin{equation}
\frac{1}{2}\zeta ^{(0)}\frac{\partial}{\partial {\bf V}}\cdot \left( {\bf
V}\boldsymbol{\mathcal{A}}\right)-\frac{1}{2}\zeta ^{(0)} \boldsymbol{\mathcal{A}}
-\left(J^{(0)}[\boldsymbol{\mathcal{A}},f^{(0)}]+J^{(0)}[f^{(0)},\boldsymbol{\mathcal{A}}]\right)
=\mathbf{A},
\label{c.2}
\end{equation}
\begin{equation}
\frac{1}{2}\zeta ^{(0)}\frac{\partial}{\partial {\bf V}}\cdot \left( {\bf
V}\boldsymbol{\mathcal{C}}\right)-n\frac{\partial \zeta ^{(0)}}{\partial n} \boldsymbol{\mathcal{A}}
-\left(J^{(0)}[\boldsymbol{\mathcal{C}},f^{(0)}]+J^{(0)}[f^{(0)},\boldsymbol{\mathcal{C}}]\right)
=\mathbf{C},
\label{c.3}
\end{equation}
where $J^{(0)}[X,Y]$ is the linearized collision operator
\begin{equation}
\label{c.4} J^{(0)}\left[{\bf v}_{1}|X,Y\right]=\chi^{(0)}\sigma^{d-1} \int
d\mathbf{v}_{2}\int d\widehat{{\boldsymbol\sigma}}\Theta (
\widehat{{\boldsymbol\sigma}}\cdot \mathbf{g})(\widehat{{\boldsymbol\sigma}} \cdot
\mathbf{g})\left[ \alpha^{-2} X(\mathbf{V}_{1}^{\prime \prime
})Y(\mathbf{V}_{2}^{\prime \prime })- X(\mathbf{V}_{1})Y(\mathbf{V}_{2})\right],
\end{equation}
and the inhomogeneous terms of the integral equations (\ref{c.2}) and (\ref{c.3}) are defined by
\begin{equation}
A_{i}\left( \mathbf{V}\right)=\frac{1}{2} V_{i}\frac{\partial}{\partial {\bf V}}\cdot \left(
\mathbf{V}f^{(0)}\right) -\frac{p}{\rho}\frac{\partial}{\partial V_{i}}
f^{(0)}+\frac{1}{2}\mathcal{ K}_{i}\left[\frac{\partial}{\partial {\bf V}}\cdot \left( \mathbf{V}
f^{(0)}\right) \right] , \label{c.5}
\end{equation}
\begin{equation}
\label{c.6} C_{i}\left( \mathbf{V}\right) =-{\bf V} f^{(0)}-m^{-1}\frac{\partial}{\partial V_{i}}
f^{(0)}\frac{\partial p}{\partial n}- \left(1+\frac{1}{2}\phi\frac{\partial \ln \chi^{(0)}}
{\partial \phi}\right)\mathcal{K}_{i}\left[f^{(0)}\right].
\end{equation}
Here, the operator $\mathcal{K}_{i}[X]$ is given by
\begin{equation}
\label{c.7} \mathcal{K}_{i}[X] =\sigma^{d}\chi^{(0)}\int d \mathbf{v}_{2}\int
d\widehat{\boldsymbol {\sigma }}\Theta (\widehat{\boldsymbol {\sigma}} \cdot
\mathbf{g})(\widehat{\boldsymbol {\sigma}}\cdot \mathbf{g}) \widehat{\sigma}_i \left[
\alpha^{-2}f^{(0)}(\mathbf{V}_{1}^{\prime \prime})X(\mathbf{V}_{2}^{\prime \prime
})+f^{(0)}(\mathbf{V}_{1})X(\mathbf{V}_{2})\right] .\nonumber\\
\end{equation}

The functions $\boldsymbol{\mathcal{A}}$ and $\boldsymbol{\mathcal{C}}$ are zero in the
first Sonine approximation \cite{GD99a,L05}. In the second Sonine approximation, these
quantities are given by
\begin{equation}
\label{a.1} \boldsymbol{\mathcal{A}}({\bf V})\to -f_{M}({\bf V})a\; {\bf S}({\bf V}),\quad
\boldsymbol{\mathcal{C}}({\bf V})\to -f_{M}({\bf V})c\; {\bf S}({\bf V}),
\end{equation}
where
\begin{equation}
\label{a.2} f_{M}({\bf V})=n \left(\frac{m}{2\pi T}\right)^{d/2}\exp\left(-\frac{mV^2}{2T}\right),
\end{equation}
and
\begin{equation}
\label{a.3} {\bf S}({\bf V})=\left(\frac{1}{2}mV^2-\frac{d+2}{2}T\right){\bf V}.
\end{equation}
Substitution of (\ref{a.1}) into Eqs.\ (\ref{c.2}) and (\ref{c.3}) gives a set of
closed equations for $a$ and $c$. Multiplication of these equations by ${\bf S}({\bf
V})$ and integration over ${\bf V}$ yields a set of algebraic equations whose solution
is \cite{GD99a,L05}
\begin{equation}
\label{a.4} a=\frac{1+3\frac{2^{d-3}}{d+2}\phi\chi^{(0)} (1+\alpha)^2(2\alpha-1)}{\nu
T^2
 (\nu_{\kappa}^*-2\zeta^{*})},
\end{equation}
\begin{equation}
\label{a.5} c=\frac{1}{nT\nu}\left(\nu_\kappa^*-\frac{3}{2}\zeta^{*}\right)^{-1}\left[
aT^2\nu \frac{\xi\zeta^{*}}{\chi^{(0)}}-3\frac{2^{d-3}}{d+2}\phi(
\chi^{(0)}+\xi)\alpha(1-\alpha^2)\right],
\end{equation}
where $\nu=n\sigma^{d-1}\sqrt{2T/m}$, $\zeta^*$ is given by (\ref{4.16}),
\begin{equation}
\xi\equiv \frac{\partial}{\partial\phi}\left(\phi \chi^{(0)}\right) \label{a.6}
\end{equation}
and
\begin{equation}
\label{a.7} \nu_\kappa^*=\frac{8}{d(d+2)}\frac{\pi^{(d-1)/2}}{\sqrt{2}\Gamma(d/2)}
\chi^{(0)}(1+\alpha)\left[\frac{d-1}{2}+\frac{3}{16}(d+8)(1-\alpha)\right].
\end{equation}
It must be remarked that all the above expressions have been obtained by neglecting
some non-Gaussian contributions to the zeroth-order distribution $f^{(0)}$. The
influence of these non-Gaussian terms is only significant for quite extreme values of
dissipation \cite{BR05}.

\section{First and second Sonine approximations for the mass flux of impurities}
\label{appA}

 In this Appendix we determine the transport coefficients $D_{0}$, $D$ and $D^T$ associated with the mass
flux in the first and second Sonine approximation. In this case, the functions
$\boldsymbol{\mathcal{A}}_0$, $\boldsymbol{\mathcal{B}}_0$ and
$\boldsymbol{\mathcal{C}}_0$ are given by Eqs.\ (\ref{4.1})--(\ref{4.3}), respectively
while $\boldsymbol{\mathcal{A}}$ and $\boldsymbol{\mathcal{C}}$ are approximated by
(\ref{a.1}). Let us start with the thermal diffusion coefficient $D^T$, which is
defined by Eq.\ (\ref{3.3}). To get it, we substitute first
$\boldsymbol{\mathcal{A}}_0$ and $\boldsymbol{\mathcal{A}}$ by their Sonine
approximations (\ref{4.1}) and (\ref{a.1}), respectively, and then we multiply the
integral equation (\ref{3.6}) by $m_0{\bf V}$ and integrate over velocity. After some
algebra, the result is
\begin{equation}
\label{a.9} (\nu_{1} -\zeta^{(0)})D^T+\frac{n_0T_0^2}{\rho}\nu_2 a_0=Z_1,
\end{equation}
where
\begin{equation}
\label{a.9.1} Z_1=-\frac{x_0T_0^2}{m}\nu_3 a-\frac{x_0pm_0}{m\rho}\left(1-\frac{\rho
T_0}{m_0p}\right)- \frac{1}{2d\rho}\int d{\bf v} m_0 V_i \mathcal{ K}_{0,i
}\left[\frac{\partial}{\partial {\bf V}}\cdot \left( \mathbf{V} f^{(0)}\right) \right],
\end{equation}
and we have introduced the collision frequencies
\begin{equation}
\label{a.10} \nu_{1}=-\frac{1}{dn_0T_0}\int d{\bf v}\,m_0{\bf V}\cdot
J_{0}^{(0)}[f_{0,M}{\bf V},f^{(0)}],
\end{equation}
\begin{equation}
\label{a.11} \nu_{2}=-\frac{1}{dn_0T_0^2} \int d{\bf v}\,m_0{\bf V}\cdot J_{0}^{(0)}[f_{0,M}{\bf
S}_0,f^{(0)}],
\end{equation}
\begin{equation}
\label{a.12} \nu_{3}=-\frac{1}{dn_0T_0^2} \int d{\bf v}\,m_0{\bf V}\cdot
J_{0}^{(0)}[f_0^{(0)},f_{M}{\bf S}].
\end{equation}
The operator $\mathcal{ K}_{0,i}[X]$ is defined in Eq.\ (\ref{3.14}). The collision
integral appearing on the right-hand side of Eq.\ (\ref{a.9.1}) involving this operator
has been evaluated in Ref.\ \cite{GHD07} with the result
\begin{equation}
\label{a.13} \frac{1}{2d\rho}\int d{\bf v} m_0 V_i \mathcal{ K}_{0,i
}\left[\frac{\partial}{\partial {\bf V}}\cdot \left( \mathbf{V} f^{(0)}\right)
\right]=-\frac{1}{2}\frac{x_0T}{m}(1+\omega)^d M_{0}\phi
 \chi_0^{(0)} (1+\alpha_0),
\end{equation}
where $\omega\equiv \sigma_0/\sigma$ and $M_{0}\equiv m_0/(m+m_0)$. If only the first
Sonine correction is retained (which means $a_0=a=0$), the solution to Eq.\ (\ref{a.9})
is
\begin{equation}
\label{a.14} D^T[1]=-x_0\left(\nu_{1} -\zeta^{(0)}\right)^{-1}\left[\frac{p
m_0}{m^2n}\left(1-\frac{\rho T_0}{m_0p}\right)-\frac{1}{2}\frac{M_{0}}{m}(1+\omega)^d
\phi
 \chi_0^{(0)} (1+\alpha_0)\right].
\end{equation}
Here, $D^T[1]$ denotes the first Sonine approximation to $D^T$. To close the
determination of $D^T$ up to the second Sonine approximation, we multiply now Eq.\
(\ref{3.6}) by ${\bf S}_0({\bf V})$ and integrate over velocity to get
\begin{equation}
\label{a.15}(\nu_{4}
-2\zeta^{(0)})a_0+\frac{\rho}{n_0T_0^2}(\nu_{5}-\zeta^{(0)})D^T=Z_2,
\end{equation}
where
\begin{equation}
\label{a.16} Z_2= -a\nu_{6}+\frac{1}{T_0}-\frac{1}{d(d+2)}\frac{m_0}{n_0T_0^3}\int
d{\bf v} S_{0,i} \mathcal{K}_{0,i}\left[\frac{\partial}{\partial {\bf V}}\cdot \left(
\mathbf{V} f^{(0)}\right) \right],
\end{equation}
\begin{equation}
\label{a.17} \nu_{4}=-\frac{2}{d(d+2)}\frac{m_0}{n_0T_0^3}\int d{\bf v}\,{\bf S}_0\cdot
J_{0}^{(0)}[f_{0,M}{\bf S}_0,f^{(0)}],
\end{equation}
\begin{equation}
\label{a.19} \nu_{5}=-\frac{2}{d(d+2)}\frac{m_0}{n_0T_0^2}\int d{\bf v}\,{\bf S}_0\cdot
J_{0}^{(0)}[f_{0,M}{\bf V},f^{(0)}],
\end{equation}
\begin{equation}
\label{a.18} \nu_{6}=-\frac{2}{d(d+2)}\frac{m_0}{n_0T_0^3}\int d{\bf v}\,{\bf S}_0\cdot
J_{0}^{(0)}[f_0^{(0)},f_{M}{\bf S}_0].
\end{equation}
The collision integral of (\ref{a.16}) involving the operator $\mathcal{K}_{0,i}$ is
given by \cite{GHD07}
\begin{eqnarray}
\label{a.20} \frac{1}{d}\int d{\bf v}& & S_{0,i}({\bf V}){\cal K}_{0,i}
\left[\frac{\partial}{\partial {\bf V}}\cdot({\bf V}f^{(0)})\right]=-\frac{1}{2}x_0\frac{n
M_{0}T^2}{m}(1+\omega)^d \chi_0^{(0)}\phi (1+\alpha_0)\left\{
\frac{M\gamma}{M_0}\left[(d+2)(M_{0}^2-1)\right.\right.\nonumber\\
& & \left.\left. +(2d-5-9\alpha_0)M_{0}M+ (d-1+3\alpha_0+6\alpha_0^2)M^2\right]+6
M^2(1+\alpha_0)^2\right\},
\end{eqnarray}
where $\gamma \equiv T_0/T$ is the temperature ratio and $M\equiv m/m_0+m$. In reduced
units and by using matrix notation, Eqs.\ (\ref{a.9}) and (\ref{a.15}) can be rewritten
as
\begin{equation}
\label{a.21} \left(
\begin{array}{cc}
\nu_{1}^*-\zeta^*&\gamma^2\nu_{2}^*\\
\frac{\nu_{5}^*-\zeta^*}{\gamma^2}&\nu_{4}^*-2\zeta^*
\end{array}
\right) \left(
\begin{array}{c}
D^{T*}\\
a_0^* \end{array} \right)= \left(
\begin{array}{c}
X_1^*-\gamma^2a^*\nu_{3}^*\\
X_2^*-a^*\nu_{6}^* \end{array} \right).
\end{equation}
Here, $\nu_{i}^*=\nu_{i}/\nu$, $D^{T*}=(m \nu/x_0T)D^T$, $a_0^*=T^2\nu a_0$,
$a^*=T^2\nu a$, and
\begin{equation}
\label{a.21.1} X_1^*=-\left(\frac{m_0 p}{m nT}-\gamma\right)+\frac{1}{2} (1+\omega)^d
M_{0}\chi_0^{(0)}\phi (1+\alpha_0),
\end{equation}
\begin{eqnarray}
\label{a.21.2} X_2^*&=&\gamma^{-1}+\frac{1}{2(d+2)} \frac{M_{0}^2}{M} (1+\omega)^d
\gamma^{-3}\chi_0^{(0)}\phi (1+\alpha_0)\nonumber\\
& & \times \left\{\frac{M
\gamma}{M_0}\left[(d+2)(M_{0}^2-1)+(2d-5-9\alpha_0)M_{0}M\right.\right.
\nonumber\\
& & \left.\left.+(d-1+3\alpha_0+6\alpha_0^2)M^2\right]+6 M^2(1+\alpha_0)^2\right\}.
\end{eqnarray}
The solution to Eq.\ (\ref{a.21}) provides the explicit expression of the second Sonine
approximation $D^{T}[2]$ to $D^{T}$. It can be written in the form (\ref{4.10}), where
the dimensionless function $H$ is
\begin{equation}
\label{a.22} H=\frac{\nu_1^*-\zeta^*}{X_1^*}
\frac{(\nu_{4}^*-2\zeta^*)(X_1^*-\gamma^2\nu_{3}^*a^*)-\gamma^2\nu_{2}^*
(X_2^*-\nu_{6}^*a^*)}{\nu_{2}^*(\zeta^*-\nu_{5}^*)+(\nu_{1}^*-\zeta^*)(\nu_{4}^*-2\zeta^*)}.
\end{equation}

The determination of the first and second Sonine approximations to the diffusion
coefficients $D_{0}$ and $D$ follows similar mathematical steps as those made before
for $D^T$. Here, only the final results will be provided. The kinetic diffusion
coefficient $D_{0}$ is obtained from the integral equation (\ref{3.7}) by substitution
of the Sonine approximation (\ref{4.2}). In matrix notation, the coefficients $D_{0}$
and $b_0$ obey the equation
\begin{equation}
\label{a.23} \left(
\begin{array}{cc}
\nu_{1}^*-\frac{1}{2}\zeta^*&\gamma^2\nu_{2}^*\\
\frac{\nu_{5}^*-\zeta^*}{\gamma^2}&\nu_{4}^*-\frac{3}{2}\zeta^*
\end{array}
\right) \left(
\begin{array}{c}
D_{0}^{*}\\
b_0^* \end{array} \right)= \left(
\begin{array}{c}
\gamma\\
0\end{array} \right),
\end{equation}
where $D_{0}^*=(m_0^2\nu/\rho T)D_{0}$ and $b_0^*=T\nu b_0$. The solution to
(\ref{a.23}) gives the second Sonine approximation to $D_{0}$. It can be written as
Eq.\ (\ref{4.8}) where the dimensionless function $F$ is
\begin{equation}
\label{a.24} F=\left[1+\frac{\nu_{2}^*(\zeta^*-\nu_{5}^*)}
{(\nu_{1}^*-\frac{1}{2}\zeta^*)(\nu_{4}^*-\frac{3}{2}\zeta^*)}\right]^{-1}.
\end{equation}

The corresponding matrix equation defining the coefficients $D$ and $c_0$ is
\begin{equation}
\label{a.26} \left(
\begin{array}{cc}
\nu_{1}^*-\frac{1}{2}\zeta^*&\gamma^2\nu_{2}^*\\
\frac{\nu_{5}^*-\zeta^*}{\gamma^2}&\nu_{4}^*-\frac{3}{2}\zeta^*
\end{array}
\right) \left(
\begin{array}{c}
D^{*}\\
c_0^* \end{array} \right)= \left(
\begin{array}{c}
Y_1^{*}-\gamma^2\nu_{3}^*c^*\\
Y_2^* -\nu_{6}^*c^*\end{array} \right),
\end{equation}
where $D^*=(m_0\nu/n_0 T)D$, and $c_0^*=nT\nu c_0$. Moreover, the inhomogeneous terms
are given by
\begin{equation}
\label{a.28} Y_1^*=\zeta^*D^{T*}\left(1+\phi\frac{\partial\ln \chi^{(0)}}{\partial
\phi}\right)-\frac{m_0}{m T}\frac{\partial p}{\partial n}+ \frac{1}{2} M_{0}\phi
(1+\alpha_0) \left(\frac{1+\theta}{\theta}\right)\frac{\partial}{\partial \phi}
\left(\frac{\mu_0}{T}\right)_{T,n_0},
\end{equation}
\begin{eqnarray}
\label{a.29} Y_2^*&=&\zeta^*a_0^*\left(1+\phi\frac{\partial\ln \chi^{(0)}}{\partial
\phi}\right)+\frac{1}{2(d+2)}\frac{M^2}{M_{0}}\phi (1+\alpha_0)\frac{\partial}{\partial \phi}
\left(\frac{\mu_0}{T}\right)_{T,n_0} \nonumber\\
& & \times \left\{ \left[(d+8)M_{0}^2+(7+2d-9\alpha_0)M_{0}M+(2+d+3\alpha_0^2-
3\alpha_0)M^2\right]\theta\right.\nonumber\\
& & +3M^2(1+\alpha_0)^2\theta^3+
\left[(d+2)M_{0}^2+(2d-5-9\alpha_0)M_{0}M+(d-1+3\alpha_0+6\alpha_0^2)
M^2\right]\theta^2\nonumber\\
& & \left.-(d+2)\theta(1+\theta)\right\},
\end{eqnarray}
where $c^*=nT\nu c$, $\theta=m_0T/mT_0$ is the mean-square velocity of the gas
particles relative to that of the intruder particle and
\begin{equation}
\label{2.29.1}
a_0^*=\frac{\gamma^{-2}(\zeta^*-\nu_{5}^*)(X_1^*-\gamma^2\nu_{3}^*a^*)+(\nu_{1}^*-\zeta^*)
(X_2^*-\nu_{6}^*a^*)}{\nu_{2}^*(\zeta^*-\nu_{5}^*)+(\nu_{1}^*-\zeta^*)(\nu_{4}^*-2\zeta^*)}.
\end{equation}
The second Sonine approximation $D[2]$ can be easily obtained from Eq.\ (\ref{a.26})
and the result can be written in the form (\ref{4.9}) where $G$ is
\begin{equation}
\label{a.30} G=\frac{\nu_{1}^*-\frac{3}{2}\zeta^*-
\gamma^2\nu_{2}^*\left(Y_2^*-\nu_{6}^*c^*\right)
\left(Y_1^{*}-\gamma^2\nu_{3}^*c^*\right)^{-1}}
{\nu_{1}^*-\frac{3}{2}\zeta^*+\nu_{2}^*\left(\zeta^*-\nu_{5}^*\right)
\left(\nu_{1}^*-\frac{1}{2}\zeta^*\right)^{-1}}.
\end{equation}

Finally, in order to get the explicit dependence of the transport coefficients on
dissipation, one still needs to compute the temperature ratio $\gamma=T_0/T$. It is
determined from the condition $\zeta_0^*=\zeta^*$, where $\zeta_0^*$ is \cite{GHD07}
\begin{equation}
\label{a.34} \zeta_0^*=\frac{4\pi^{(d-1)/2}}{d\Gamma\left(\frac{d}{2}\right)}\left(\frac{\overline{\sigma}}{\sigma}
\right)^{d-1}\chi_0^{(0)}M\left(\frac{1+\theta}{\theta}\right)^{1/2}(1+\alpha_0)
\left[1-\frac{M}{2}(1+\theta)(1+\alpha_0)\right].
\end{equation}

\section{Collision integrals}
\label{appB}

In this Appendix we obtain the expressions for the collision frequencies $\nu_{i}^*$.
Except $\nu_{2}^*$ and $\nu_{3}^*$, the other quantities were already determined
\cite{GHD07} for arbitrary composition. For the sake of completeness, we display now
their explicit forms in the tracer limit ($x_1\to 0$). They are given by
\begin{equation}
\label{b.1}
\nu_{1}^*=\frac{2\pi^{(d-1)/2}}{d\Gamma\left(\frac{d}{2}\right)}\left(\frac{\overline{\sigma}}{\sigma}
\right)^{d-1}\chi_0^{(0)}M(1+\alpha_0) \left(\frac{1+\theta}{\theta}\right)^{1/2},
\end{equation}
\begin{equation}
\label{b.2} \nu_{4}^*=\frac{\pi^{(d-1)/2}}
{d(d+2)\Gamma\left(\frac{d}{2}\right)}\left(\frac{\overline{\sigma}}{\sigma}\right)^{d-1}
\chi_0^{(0)}M(1+\alpha_0)\left(\frac{\theta}{1+\theta}\right)^{3/2}
\left[A-(d+2)\frac{1+\theta}{\theta} B\right],
\end{equation}
\begin{equation}
\label{b.3} \nu_{5}^*=\frac{2\pi^{(d-1)/2}}
{d(d+2)\Gamma\left(\frac{d}{2}\right)}\left(\frac{\overline{\sigma}}{\sigma}\right)^{d-1}
\chi_0^{(0)}M(1+\alpha_0)\left(\frac{\theta}{1+\theta}\right)^{1/2}B,
\end{equation}
\begin{equation}
\label{b.4} \nu_{6}^*=-\frac{\pi^{(d-1)/2}}
{d(d+2)\Gamma\left(\frac{d}{2}\right)}\left(\frac{\overline{\sigma}}{\sigma}\right)^{d-1}
\chi_0^{(0)}\frac{M^2}{M_{0}}(1+\alpha_0)\left(\frac{\theta}{1+\theta}\right)^{3/2}
\left[C+(d+2)\frac{1+\theta}{\theta} D\right],
\end{equation}
where
\begin{eqnarray}
\label{b.5} A&=&
 2M^2\left(\frac{1+\theta}{\theta}\right)^{2}
\left(2\alpha_0^{2}-\frac{d+3}{2}\alpha_0+d+1\right)
\left[d+5+(d+2)\theta\right]\nonumber\\
& & -M(1+\theta) \left\{\lambda\theta^{-2}[(d+5)+(d+2)\theta][(11+d)\alpha_0
-5d-7]\right.\nonumber\\
& & \left. -\theta^{-1}[20+d(15-7\alpha_0)+d^2(1-\alpha_0)-28\alpha_0] -(d+2)^2(1-\alpha_0)\right\}
\nonumber\\
& & +3(d+3)\lambda^2\theta^{-2}[d+5+(d+2)\theta]+ 2\lambda\theta^{-1}[24+11d+d^2+(d+2)^2\theta]
\nonumber\\
& & +(d+2)\theta^{-1} [d+3+(d+8)\theta]-(d+2)(1+\theta)\theta^{-2}
[d+3+(d+2)\theta],\nonumber\\
\end{eqnarray}
\begin{eqnarray}
\label{b.6} B&=& (d+2)(1+2\lambda)+M(1+\theta)\left\{(d+2)(1-\alpha_0)
-[(11+d)\alpha_0-5d-7]\lambda\theta^{-1}\right\}\nonumber\\
& & +3(d+3)\lambda^2\theta^{-1}+2M^2\left(2\alpha_0^{2}-\frac{d+3}{2}\alpha
_{12}+d+1\right)\theta^{-1}(1+\theta)^2\nonumber\\
& & - (d+2)\theta^{-1}(1+\theta),
\end{eqnarray}
\begin{eqnarray}
\label{b.7} C&=&
 2M^2(1+\theta)^2
\left(2\alpha_0^{2}-\frac{d+3}{2}\alpha_0+d+1\right)
\left[d+2+(d+5)\theta\right]\nonumber\\
& & -M(1+\theta) \left\{\lambda[d+2+(d+5)\theta][(11+d)\alpha_0
-5d-7]\right.\nonumber\\
& & \left. +\theta[20+d(15-7\alpha_0)+d^2(1-\alpha_0)-28\alpha_0] +(d+2)^2(1-\alpha_0)\right\}
\nonumber\\
& & +3(d+3)\lambda^2[d+2+(d+5)\theta]- 2\lambda[(d+2)^2+(24+11d+d^2)\theta]
\nonumber\\
& & +(d+2)\theta [d+8+(d+3)\theta]-(d+2)(1+\theta)
[d+2+(d+3)\theta], \nonumber\\
\end{eqnarray}
\begin{eqnarray}
\label{b.5.1} D&=& (d+2)(2\lambda-\theta)+M(1+\theta)\left\{(d+2)(1-\alpha_0)
+[(11+d)\alpha_0-5d-7]\lambda\right\}\nonumber\\
& & -3(d+3)\lambda^2-2M^2\left(2\alpha_0^{2}-\frac{d+3}{2}\alpha _{12}+d+1\right)
(1+\theta)^2+(d+2)(1+\theta),
\end{eqnarray}
Here, $\lambda=M_{0}(1-\gamma^{-1})$. It must be noticed that Eqs.\ (\ref{b.1})--(\ref{b.5.1})
have been obtained by taking Maxwellian distributions for the reference homogeneous
cooling state distributions $f^{(0)}$ and $f_0^{(0)}$.

It only remains to evaluate the collision integrals defining the
collision frequencies $\nu_{2}^*$ and $\nu_{3}^*$. To compute them, we use the property
\begin{eqnarray}
\int d{\bf v}_{1}h({\bf v}_{1})J_{0}^{(0)}\left[ {\bf V}_{1}|F,G\right] &=&
\chi_0^{(0)}\overline{\sigma}^{d-1} \int \,d{\bf v}_{1}\,\int \,d{\bf v}_{2}\;F(
{\bf v}_{1})G({\bf v}_{2})  \nonumber \\
&&\times \int d\widehat{\boldsymbol {\sigma }}\,\Theta (\widehat{\boldsymbol {\sigma}} \cdot {\bf
g})(\widehat{\boldsymbol {\sigma }}\cdot {\bf g})\,\left[ h( {\bf V}_{1}^{^{\prime}})-h({\bf V}_{1})\right] \;,
\label{b.8}
\end{eqnarray}
with
\begin{equation}
{\bf V}_{1}'={\bf V}_{1}-M(1+\alpha _{0})( \widehat{\boldsymbol {\sigma }}\cdot {\bf
g})\widehat{\boldsymbol {\sigma}}\;. \label{b.9}
\end{equation}
Use of this property in Eq.\ (\ref{a.11}) gives
\begin{equation}
\label{b.9.1} \nu_{2}^*=\frac{2\pi^{(d-1)/2}}
{d\Gamma\left(\frac{d+3}{2}\right)}\left(\frac{\overline{\sigma}}{\sigma}\right)^{d-1}\chi_0^{(0)}
M(1+\alpha_0) \theta^{1+\frac{d}{2}}\int\;d{\bf c}_1\int\;d{\bf c}_2\; {\bf
x}e^{-\theta c_1^2-c_2^2}\left(\theta c_1^2-\frac{d+2}{2}\right) ({\bf x}\cdot {\bf
c}_1),
\end{equation}
where ${\bf c}_i={\bf v}_i/v_0$, ${\bf x}={\bf g}/v_0$, and we have taken the Maxwellian approximation
(\ref{a.2}) for $f^{(0)}$. The integrals appearing in (\ref{b.9}) can be evaluated by the change of variables
$\{{\bf c}_1, {\bf c}_2\}\to \{{\bf x}, {\bf y}\}$, where ${\bf y}=\theta {\bf c}_1+{\bf c}_2$, the Jacobian being  $(1+\theta)^{-d}$. With this change the integrals can be easily performed and the final result is
\begin{equation}
\label{b.10} \nu_{2}^*=\frac{\pi^{(d-1)/2}}
{d\Gamma\left(\frac{d}{2}\right)}\left(\frac{\overline{\sigma}}{\sigma}\right)^{d-1}\chi_0^{(0)}
M(1+\alpha_0)[\theta(1+\theta)]^{-1/2}.
\end{equation}
Similarly, the reduced collision frequency $\nu_{3}^*$ is given by
\begin{equation}
\label{b.11} \nu_{3}^*=-\frac{\pi^{(d-1)/2}}
{d\Gamma\left(\frac{d}{2}\right)}\left(\frac{\overline{\sigma}}{\sigma}\right)^{d-1}\chi_0^{(0)}
\frac{M^2}{M_{0}}(1+\alpha_0) \theta^{3/2}(1+\theta)^{-1/2}.
\end{equation}

\section{Modified Sonine approximation }
\label{appD}

The expression for the kinetic diffusion coefficient $D_0^*$ derived from a modified version of the first Sonine approximation recently proposed \cite{GSM07,GVM09} is displayed in this Appendix. This coefficient is given by
\begin{equation}
\label{d1}
D_0^*=\frac{\gamma}{\nu_D^*-\frac{1}{2}\zeta^*}.
\end{equation}
Here,
\begin{equation}
\label{d2}
\zeta^{*}=\frac{\sqrt{2}\pi^{(d-1)/2}}{d\Gamma(d/2)}\chi^{(0)}(1-\alpha^2)\left(1+\frac{3}{32}e\right)
\end{equation}
where $e$ is the fourth cumulant of the gas distribution function $f^{(0)}$. It measures the departure of $f^{(0)}$ from its Maxwellian form and its expression is \cite{NE98}
\begin{equation}
\label{d3}
e(\alpha)=\frac{32(1-\alpha)(1-2\alpha^2)}{9+24d-(41-d)\alpha+30\alpha^2(1-\alpha)}.
\end{equation}
The (reduced) collision frequency $\nu_D^*$ is given by \cite{GVM09}
\begin{equation}
\label{d4}
\nu_{D}^*=\frac{2\pi^{(d-1)/2}}{d\Gamma\left(\frac{d}{2}\right)}\left(\frac{\overline{\sigma}}{\sigma}
\right)^{d-1}\chi_0^{(0)}M(1+\alpha_0) \left(\frac{1+\theta}{\theta}\right)^{1/2}\left[1+\frac{1}{16}\frac{(3+4\theta)e_{0}-\theta^2e}{(1+\theta)^2}\right],
\end{equation}
where $e_0$ is the corresponding fourth cumulant for the distribution function $f_0^{(0)}$ of the intruder. Its explicit form can be found in the Appendix of Ref.\ \cite{G08a}.


\begin{thebibliography} {99}
\bibitem{DSMCnote}Please notice that using the same number of simulated particles for both species does not imply that their densities are equal: the very small concentration limit for the tracer particles is implied by the form of the two kinetic equations that are simulated.
\end{thebibliography}
\end{document}